\documentclass[twocolumn]{aastex631}
\usepackage[version=4]{mhchem}

\newcommand{\rj}{$R_\mathrm{Jup}$}
\newcommand{\mj}{$M_\mathrm{Jup}$}

\shorttitle{Complex Organic Molecules in the Jovian Circumplanetary Disk}
\shortauthors{Mousis et al.}
\graphicspath{{./}{figures/}}

\begin{document}

\title{Formation and Survival of Complex Organic Molecules in the Jovian Circumplanetary Disk}

\correspondingauthor{Olivier Mousis}
\email{olivier.mousis@swri.org}

\author[0000-0001-5323-6453]{Olivier Mousis}
\affiliation{Solar System Science and Exploration Division, Southwest Research Institute, 1301 Walnut St, Ste 400, Boulder, CO, USA}
\author{Clément Petetin}
\affiliation{Groupe de Spectrométrie Moléculaire et Atmosphérique (GSMA), Université de Reims Champagne-Ardenne, CNRS, 51687 Reims cedex, France}
\author[0000-0002-8719-7867]{Tom Benest Couzinou}
\affiliation{Aix-Marseille Universit\'e, CNRS, CNES, Institut Origines, LAM, Marseille, France}
\author[0000-0002-3289-2432]{Antoine Schneeberger}
\affiliation{Astronomy \& Astrophysics Section, School of Cosmic Physics, Dublin Institute for Advanced Studies, 31 Fitzwilliam Place, Dublin D02 XF86, Ireland}
\author[0000-0002-3289-2432]{Yannis Bennacer}
\affiliation{Aix-Marseille Universit\'e, CNRS, CNES, Institut Origines, LAM, Marseille, France}



\begin{abstract}
Europa, Ganymede, and Callisto are key targets in the search for habitability due to the potential presence of subsurface oceans. Detecting complex organic molecules (COMs), essential for prebiotic chemistry, is crucial to assessing their potential. Though COMs remain undetected on these moons, ESA's JUICE and NASA's Europa Clipper missions aim to fill this gap with their science payloads. This study explores the formation and transport of COMs within Jupiter’s circumplanetary disk (CPD), a critical environment for the formation of the Galilean moons. Using a time--dependent model that couples the evolving CPD structure with the dynamics of icy particles of varying sizes and release times, we assess two primary COM formation pathways: thermal processing of ices and UV photochemistry. The results indicate that heating, particularly of NH$_3$:CO$_2$ ices, occurs efficiently before substantial irradiation, making it the dominant pathway for COM formation in the Jovian CPD. However, the relative efficiencies of these two processes are governed by particle density, disk viscosity, accretion rate, and UV flux, which collectively determine drift timescales and exposure to favorable thermodynamic environments. Existing models indicate that Europa’s accretion was relatively cold and prolonged, possibly allowing some COMs to survive incorporation, whereas Ganymede and Callisto likely formed under even cooler conditions conducive to preserving COM--rich material. These results highlight the potential inheritance of complex organics by the Galilean moons and offer a framework for interpreting upcoming compositional data from JUICE and Europa Clipper.

\end{abstract}



\section{Introduction} 
\label{sec:sec1}

Europa, Ganymede, and Callisto are believed to have subsurface oceans beneath their icy crusts \citep{Ki00}, positioning them as key targets in the search for habitable environments within the Solar System \citep{tocard13,Saur15,Vance23}. It is critical to constrain the internal composition of these moons, particularly the presence of complex organic molecules (COMs), which are composed of carbon, hydrogen, oxygen, and often nitrogen, in order to evaluate their potential habitability. COMs are fundamental precursors to biomolecules, such as amino acids and nucleobases, that play a key role in prebiotic chemistry. The presence of COMs indicates ongoing organic chemical processes and suggests that these moons may contain the essential ingredients for life: liquid water, energy sources, and organic compounds \citep{Mc08,Ha09}.

There is still a lack of direct evidence for COMs in Europa, Ganymede, and Callisto. However, the upcoming ESA Jupiter Icy Moons Explorer (JUICE) \citep{tocard13} and NASA Europa Clipper missions \citep{Bec24,papp24} are likely to address this gap. Both missions will carry advanced instruments, including infrared spectrometers, submillimeter sensors, and mass spectrometers. These instruments are designed to analyze the surface and exospheric compositions of these icy moons with unparalleled precision \citep{Wu18,Ha23,Wa24,Po24}. These observations are expected to provide valuable information about the chemical environments of the moons, including the presence and distribution of complex organic compounds, salts, and volatile ices. All of these are key indicators of potential habitability.

The formation and transport of COMs in protoplanetary disks (PPDs) have been extensively investigated \citep{Wa14,Ben24,Benest_Couzinou_2025}. In contrast, comparatively little attention has been devoted to the synthesis, alteration, or survival of COMs within circumplanetary disks (CPDs), despite these being the environments in which the satellites of gas giants are thought to form \citep{canup2002, canup2006, Ronnet2017, anderson2021}. Although COMs can form efficiently in the protosolar nebula (PSN), their delivery to the forming Galilean moons is severely constrained. Inward transport toward Jupiter’s orbit is inefficient and strongly size-dependent, while gap formation, irradiation, and thermal processing during transfer to the CPD likely destroy or chemically alter a large fraction of COMs. Consequently, only a limited fraction of PSN-formed COMs is expected to survive and be incorporated into moon-forming material \citep{Ben25}. By comparison, the physical and chemical conditions in CPDs differ markedly from those in PPDs, opening the possibility of distinct chemical pathways. While elevated temperatures in CPDs can further degrade COMs inherited from the PPD, they may also promote in situ chemical processing of icy particles delivered from the PPD, enabling the formation of new COMs directly within the CPD.

The formation and transport of COMs in protoplanetary disks (PPDs) have been extensively investigated \citep{Wa14,Ben24,Benest_Couzinou_2025}. In contrast, comparatively little attention has been devoted to the synthesis, alteration, or survival of COMs within circumplanetary disks (CPDs), despite these being the environments in which the satellites of gas giants are thought to form \citep{canup2002, canup2006, Ronnet2017, anderson2021}. Although COMs can form efficiently in the protosolar nebula (PSN), their delivery to the forming Galilean moons is severely constrained. Inward transport toward Jupiter’s orbit is inefficient and strongly size-dependent, while gap formation, irradiation, and thermal processing during transfer to the CPD likely destroy or chemically alter a large fraction of COMs. Consequently, only a limited fraction of PSN-formed COMs is expected to survive and be incorporated into moon-forming material \citep{Ben25}. By comparison, the physical and chemical conditions in CPDs differ markedly from those in PPDs, opening the possibility of distinct chemical pathways. While elevated temperatures in CPDs can further degrade COMs inherited from the PPD, they may also promote in situ chemical processing of icy particles delivered from the PPD, enabling the formation of new COMs directly within the CPD.

In this study, we investigate the formation and transport of COMs within a CPD, using the model developed by \citet{Schneeberger2024}, in which the accretion rate reflects the final stages of Jupiter’s growth. This scenario leads the CPD to evolve from a hot to a cold state, encompassing a broad spectrum of thermodynamic conditions. We employ a time-dependent model that couples the evolving disk structure with the dynamics of icy particles of various sizes, released at different stages of the evolution of the CPD. We explore two COM formation pathways, both of which are grounded in laboratory experiments: (1) the thermal processing of ices within specific temperature ranges \citep{bossa2008}, and (2) the irradiation of particles by UV photons \citep{bossa2008,tenelanda2022}.
 
Section \ref{sec:sec2} details the CPD model and the framework for simulating particle transport used in this study. It also explains how the thermodynamic conditions necessary for COM formation via thermal processing and irradiation are incorporated into the model. Section \ref{sec:sec3} presents the resulting COM formation pathways within the evolving CPD environment. Finally, Section \ref{sec:sec4} explores how sensitive our results are to a larger set of parameters. This section also discusses the broader implications of our findings for satellite formation and the potential for capturing and preserving COMs during this period.

\section{Model} 
\label{sec:sec2}

\subsection{Jovian Circumplanetary Disk}
\label{sec:sec2.1}

The CPD model used in this study is axisymmetric and assumes hydrostatic equilibrium. We adopt the two-dimensional framework presented by \cite{Schneeberger2024}, which builds upon the models developed by \cite{makalkin2014} and \cite{heller2015}. This framework incorporates the viscous accretion disk prescription from \cite{canup2002}. The outer boundary of the CPD is assumed to be located at 133~\rj, which corresponds to one-fifth of Jupiter's Hill radius. The temperature and density profiles within the CPD are governed by its time-dependent accretion rate, which is parameterized following the formulation of \cite{sasaki2010}:

\begin{equation}  
\dot{M}(t) = \dot{M}_0 \exp{\left(-\frac{t}{\tau}\right)} M_\text{J}/\text{yr},  
\label{eq:sasaki_accretion}  
\end{equation}  

\noindent where $\dot{M}(t)$ represents the disk accretion rate at time $t$, and $\dot{M}_0$ corresponds to the accretion rate at the end of Jupiter's gas runaway accretion phase, when the planet reached 95\% of its current mass. The variable $\tau$ is the characteristic depletion timescale of the CPD, estimated to range from 20 kyr to 1 Myr \citep{,sasaki2010,mordasini2013}. Additionally, the CPD's thermodynamic conditions are influenced by Jupiter's mass, radius, and surface temperature. In our model, Jupiter's mass is fixed at 95$\%$ of its present value, while its radius is assumed to be 10$\%$ larger, reflecting the planet's inflated state and elevated surface temperature of approximately 2000 K \citep{mordasini2013}.  

The properties of the CPD also depend on the viscosity $\nu$, which is expressed as  

\begin{equation}  
\nu(r,z) = \alpha \frac{C_s^2(r,z)}{\Omega_\text{K}(r)},  
\end{equation}  

\noindent where $C_s(r,z)$ is the speed of sound at a given planet-centered radius, $r$, and height, $z$, above the disk mid-plane. $\Omega_\text{K}(r)$ represents the Keplerian angular frequency at radius $r$ from the planet. The constant $\alpha$, which typically ranges from $10^{-4}$ to $10^{-2}$, scales the strength of the viscosity in the CPD. This parameter is constrained by both observations and theoretical models of protoplanetary disks \citep{shakura1973, lynden-bell1974, villenave2022}.  

The structure of the CPD also depends on the location of its centrifugal radius, $R_c$. Within this radius, the CPD's gas drifts inward toward the planet, while beyond this radius it flows outward, merging with the meridional flow within the planet's Hill Sphere \citep{tanigawa2012,morbidelli2014,szulagyi2014}. In gas-starved CPD models, $R_c$ is typically assumed to be farther than the present location of Callisto, which is at 26.9~\rj~from Jupiter \citep{sasaki2010}. 

The material forming the CPD is accreted from the protoplanetary disk, which has an average molecular mass $\mu_\text{g}$ of $2.341 \times 10^{-3}$ kg.mol$^{-1}$ \citep{aguichine2022} and a metallicity of $\frac{Z}{H} = 2.45 \times 10^{-2}$ \citep{Lodders2019}. However, simulations suggest that the CPD can only accrete dust located at high altitudes within the PPD. Therefore, we assume that the dust--to--gas ratio in the accreted material is only one tenth of that in the protoplanetary disk \citep{lambrechts2012, zhu2012}. In the CPD model this results in a metallicity of $2.45 \times 10^{-3}$.

\begin{figure*}
\centering
\includegraphics[width=\linewidth]{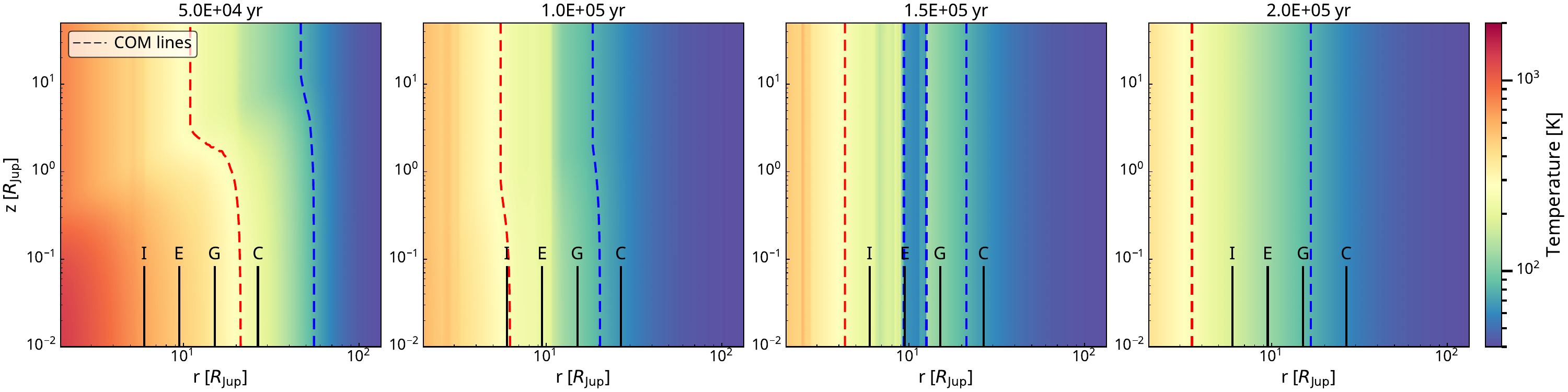}
\caption{Temperature profiles of the CPD at $t$ = 50, 100, 150, and 200 kyr of its evolution. The formation of COMs by thermal processing occurs in the temperature range from 80 K (blue dashed line) to 260 K (red dashed line). At 150 kyr of evolution, a cold region appears in the model, with temperatures too low to support COM formation by thermal processing.}
\label{fig:2D_profile}
\end{figure*}

Figure \ref{fig:2D_profile} shows the temperature and pressure profiles of a CPD model, assuming $\dot{M}_0 = 6.6 \times 10^{-6}$ \mj.yr$^{-1}$, $\tau = 2 \times 10^{4}$ yr, $\alpha = 10^{-3}$, and $R_c = 50$~\rj, as derived by \cite{mordasini2013}. This set of parameters constitutes our nominal model in the following analysis. The figure highlights a peculiar feature of the model: radiative transfer calculations produce ``shadows'' that locally cool the CPD down to 100~K below the ambient gas temperature. With this set of parameters, the disk has a short lifetime, depleting in less than 200 kyr, and transitions from being massive and hot to cold and light. This particular choice of parameters accounts for the rapid opening of a gap within the PPD at Jupiter’s location, which halts the accretion of matter \citep{mordasini2013}.

\subsection{Transport of particles}
\label{sec:sec2.2}

We use a particle transport model based on \cite{Cies10,Cies11} and \cite{Ben24}, which calculates the radial and vertical evolution of individual particles in a protoplanetary disk following a Lagrangian approach. We adapted this model to the CPD environment. The vertical evolution of particles is based on the following advection diffusion equation \citep{Du95,Gail2001,Cies10}: 
    
\begin{equation} \label{eq:adv,diff}
\frac{\partial \rho_\mathrm{s}}{\partial t} = \frac{\partial}{\partial z} \left( \rho_\mathrm{g} D \frac{\partial \frac{\rho_i}{\rho_\mathrm{g}}}{\partial z} \right) - \frac{\partial}{\partial z}\left(\rho_\mathrm{s} v_z \right),  
\end{equation}

\noindent which depends on the density of the material $\rho_\mathrm{s}$, the density of the gas $\rho_\mathrm{g}$, the diffusivity $D = \frac{\nu}{1+\text{St}^2}$, the Stokes number $\text{St}$, and the terminal velocity of the particles $v_z$ \citep{Cuzzi_Weidenschilling_2006,Cies10}. The radial evolution of the particles is analogous to this expression, but along the $x$ and $y$ cartesian axes. The vertical velocity $v_{\mathrm{eff},z}$ derived from this expression is used to compute the vertical position of the particles \citep{Cies10}:

\begin{equation} \label{eq:z_i}
z_i = z_{i-1} + v_{\mathrm{eff},z} \delta t + R_1 \left [ \frac{2}{\sigma^2}D(z)\delta t \right ]^{\frac{1}{2}}
.\end{equation}

\noindent This expression computes the vertical position $z_i$ after a time step $\delta t = 1 / \Omega_K$, given an initial vertical position $z_{i-1}$. The random motion induced by the gas turbulence is parameterized by the random number $R_1 \in [-1;1] $ and its distribution variance $\sigma^2$ (= 1/3 for a uniform distribution, \cite{Visser1997}). $v_{\textup{eff},z}$ is the vertical velocity of the particle and the sum of the following three terms \citep{Cies10}:

\begin{equation} \label{eq:v_eff,z}
v_{\mathrm{eff},z} = v_z + \frac{D}{\rho_\mathrm{g}} \frac{\partial \rho_\mathrm{g}}{\partial z} + \frac{\partial D}{\partial z},
\end{equation}

\noindent where $v_z = -t_s \Omega_K^2 z$ is the terminal velocity. The second term considers the vertical gradient of the density in the disk, and the last term considers the vertical variation of the diffusion coefficient \citep{Cies10,Ronnet2017,Ben24}. The stopping time $t_\mathrm{s}$ is calculated in the Epstein regime with the following expression \citep{Perets2011,Mousis2018}: 

\begin{equation} \label{eq:t_s}
t_\mathrm{s} = \frac{\rho_\mathrm{s}}{\rho_\mathrm{g} } \frac{R_\mathrm{s}}{v_\mathrm{th}},
\end{equation}

\noindent where $\rho_\mathrm{g}$ is the density of the gas, $R_\mathrm{s}$ is the size of a particle, and $v_\mathrm{th} = \sqrt{8/\pi} c_s$ is the thermal velocity.

The radial trajectory of the particles is computed with a similar equation, along the $x$ and $y$ axes (with $r^2 = x^2 + y^2$):

\begin{equation} \label{eq:x_i}
x_i = x_{i-1} + v_{\mathrm{eff},x} \delta t + R_1 \left [ \frac{2}{\sigma^2}D(r)\delta t \right ]^{\frac{1}{2}}
,\end{equation}

\noindent where $x_i$ and $x_{i-1}$ are the horizontal positions of the particle after and before a time step $\delta t$, respectively. Because the $y$-axis equations are the same, due to axisymmetry, only the $x$-axis equations are described here. The horizontal velocity $v_{\mathrm{eff},x}$ is the sum of three terms \citep{Cies11}:

\begin{equation} \label{eq:v_eff,x}
v_{\mathrm{eff},x} = v_r \frac{x_{i-1}}{r_{i-1}} + \frac{D}{\rho_\mathrm{g}} \frac{\partial \rho_\mathrm{g}}{\partial r} \frac{x_{i-1}}{r_{i-1}} + \frac{\partial D}{\partial r} \frac{x_{i-1}}{r_{i-1}}
.\end{equation}

\noindent Here, the last two terms are similar to those in the vertical velocity equation. The first term $v_r$ is the sum of the velocity of the gas and the radial drift of the particle. It depends on the Stokes number ($\text{St}=t_s \Omega_K$) of the particle, so small particles are less sensitive to radial drift and more coupled to the gas, and vice versa. Its expression is \citep{Birnstiel2012}:

\begin{equation} 
\label{eq:v_r}
v_r = -\frac{ 1}{1 + \mathrm{St}^2} \frac{\dot{M}}{2 \pi \Sigma r} \delta_{r_c} + \frac{2\mathrm{St} }{1+\mathrm{St}^2}\frac{c_s^2}{r\Omega_K} \frac{\mathrm{dln}P}{\mathrm{dln}r},
\end{equation}

\noindent where $\delta_{r_c}$ is a term with a value of 1 if $r>r_c$ and -1 otherwise, and $\Sigma$ is the surface density in the CPD. It accounts for the direction of the gas velocity: the gas diffuses outward in regions farther than the centrifugal radius and inward in the inner parts of the CPD. 

\begin{figure}[!ht]
\center
\includegraphics[angle=0,width=8cm]{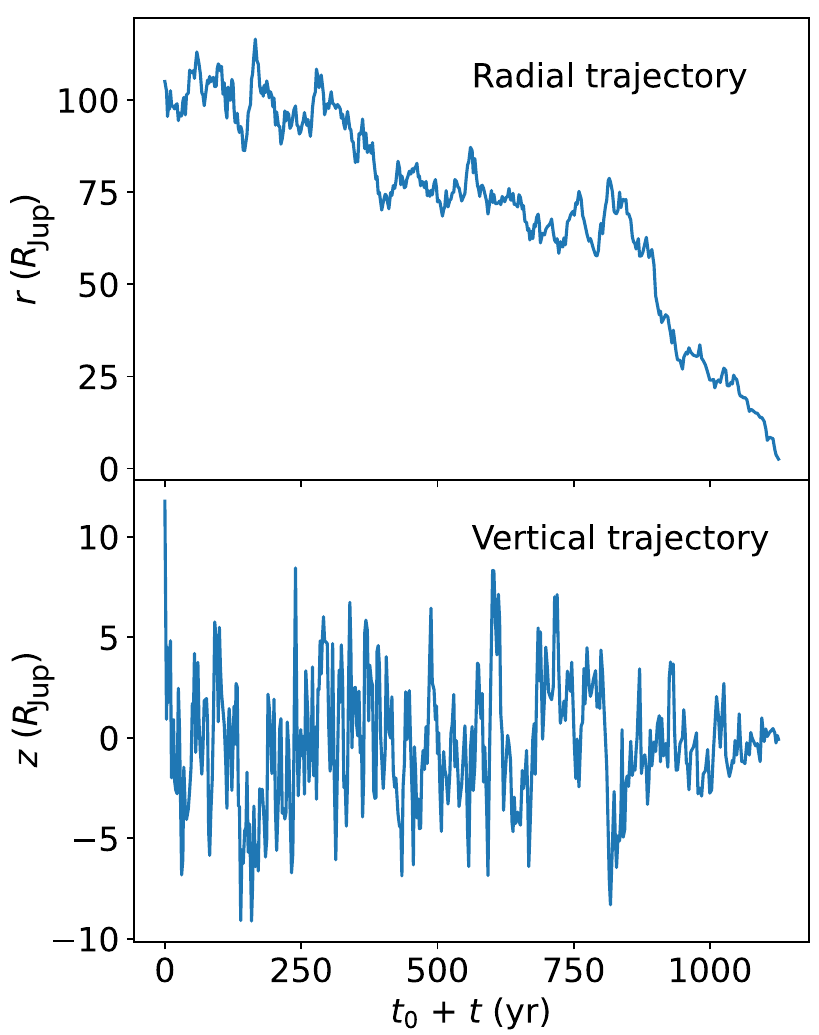}
\caption{Radial and vertical trajectory of a 100 $\mu$m particle released at one scale height above the midplane of the CPD at the distance of 115 $R_{\rm Jup}$, with an initial release time $t_{\rm 0}$~=~100 kyr, in the case of our nominal model.}
\label{fig:part}
\end{figure}

Figure~\ref{fig:part} presents the trajectory of a 100~$\mu$m particle with a density $\rho_\mathrm{s}$ of 1 g.cm$^{-3}$, released from one scale height ($H = \frac{C_s}{\Omega_K}$, corresponding to $\sim$13.8~$R_{\rm Jup}$) above the CPD midplane at a radial distance of 115~$R_{\rm Jup}$. The particle is introduced at $t_{\rm 0}$~=~100 kyr, which denotes the time elapsed since the formation of the CPD in our nominal model. After its release, the particle reaches Jupiter and stabilizes its vertical motion near the midplane in just over one thousand years.

\subsection{COM formation in the Jovian Circumplanetary Disk}
\label{sec:sec2.3}

In this study, we investigate how two distinct processes, namely irradiation and thermal processing, affect NH$_3$:CO$_2$ and CH$_3$OH ices carried by particles through the Jovian CPD. Laboratory experiments by \cite{tenelanda2022} show that COM formation in UV-irradiated pure CH$_3$OH ice depends on UV fluence and temperature, with family-dependent onset behaviors rather than a single sharp threshold. Under far-UV irradiation dominated by Lyman-$\alpha$ photons, several COMs (notably C$_2$--C$_5$ esters and ketones) are already produced at low fluence ($\sim 9\times10^{15}$ photons~cm$^{-2}$ at 20~K), while increasing fluence progressively favors heavier species through radical--radical recombination. The maximum fluence explored, $8.64\times10^{17}$ photons~cm$^{-2}$, corresponds to 24~h of irradiation at a flux of order $10^{13}$ photons~cm$^{-2}$~s$^{-1}$ and marks the upper bound of the laboratory conditions rather than a formation threshold. Similarly, \cite{bossa2008} demonstrate that NH$_3$:CO$_2$ ices generate COMs under UV irradiation, with a threshold dose of $4.32 \times 10^{19}$ photons cm$^{-2}$. These irradiation thresholds are critical parameters in our analysis. Here, the irradiation of ices is modeled as the accumulation of UV radiation within the disk, varying with the particles' position \citep{Ben24,Benest_Couzinou_2025}. The CPD is exposed to UV interstellar radiation, assumed to be incident perpendicularly to the disk plane. The UV flux, $F(r,z)$, depends on the radial ($r$) and vertical ($z$) positions within the disk and is expressed as \citep{Cies10, Ciesla_Sandford_2012, Ben24}:

\begin{equation}
F(r,z) = F_0 e^{-\tau(r,z)},
\end{equation}  

\noindent where the optical depth, $\tau(r,z)$, is expressed as  

\begin{equation} 
\tau(r,z) = \int_{|z|}^\infty \rho_{\mathrm{g}}(r,z) \kappa \, \mathrm{d}z.
\end{equation}  

\noindent The initial incident UV flux, $F_0$, is set to $10^8$ photons cm$^{-2}$ s$^{-1}$ (equivalent to G$_0$=1 in Habing units). The opacity, $\kappa$, is assumed to correspond to the frequency-averaged mean Rosseland opacity, which is the one used in the underlying CPD model.   \citep{Bell_Lin_1994, Aguichine2020, Schneeberger2023}. It is important to emphasize that, in our simulations, COM formation by irradiation of CH$_3$OH--bearing grains competes with their sublimation during inward drift in the PSN. We adopt a sublimation temperature of 105 K for CH$_3$OH, representative of PSN conditions \citep{Mousis2009}. Because these grains are small and porous, we treat sublimation as effectively instantaneous once this temperature is reached. Consequently, COM formation can only proceed if the grains accumulate the required irradiation dose before they sublimate.

In addition to UV irradiation, \cite{bossa2008} experimentally find that thermal heating drives the formation of COMs in NH$_3$:CO$_2$ ices at temperatures above 80 K. Specifically, the reactants begin to react at temperatures above 80 K and are completely consumed by 130 K. Two COMs are formed: ammonium carbamate, [NH$_2$COO$^-$][NH$_4^+$] (C), and carbamic acid, NH$_2$COOH (D), with a 1:1 ratio observed at 140 K. Both species sublimate between 230 and 260 K: C decomposes into CO$_2$ and NH$_3$ vapors at 230 K, while D partially decomposes during desorption at 260 K. Consequently, we identify the temperature range of 80–-260 K as the zone where COMs form via thermal heating and remain stable within the Jovian CPD.

\section{Results} 
\label{sec:sec3}

Simulations of particle trajectories have been performed for various particle sizes, disk parameters, and release epochs $t_{\rm 0}$ during CPD evolution. In all cases, it is assumed that the particles have a uniform density of 1~g.cm$^{-3}$. Each simulation tracks the trajectories of 800 particles of identical size, released from one scale height above the CPD midplane, and uniformly distributed across 14 release points spaced every 10 $R_{\rm Jup}$ from 5 to 135 $R_{\rm Jup}$. 

\begin{figure*}[h]
\begin{center}
\includegraphics[angle=0,width=8cm]{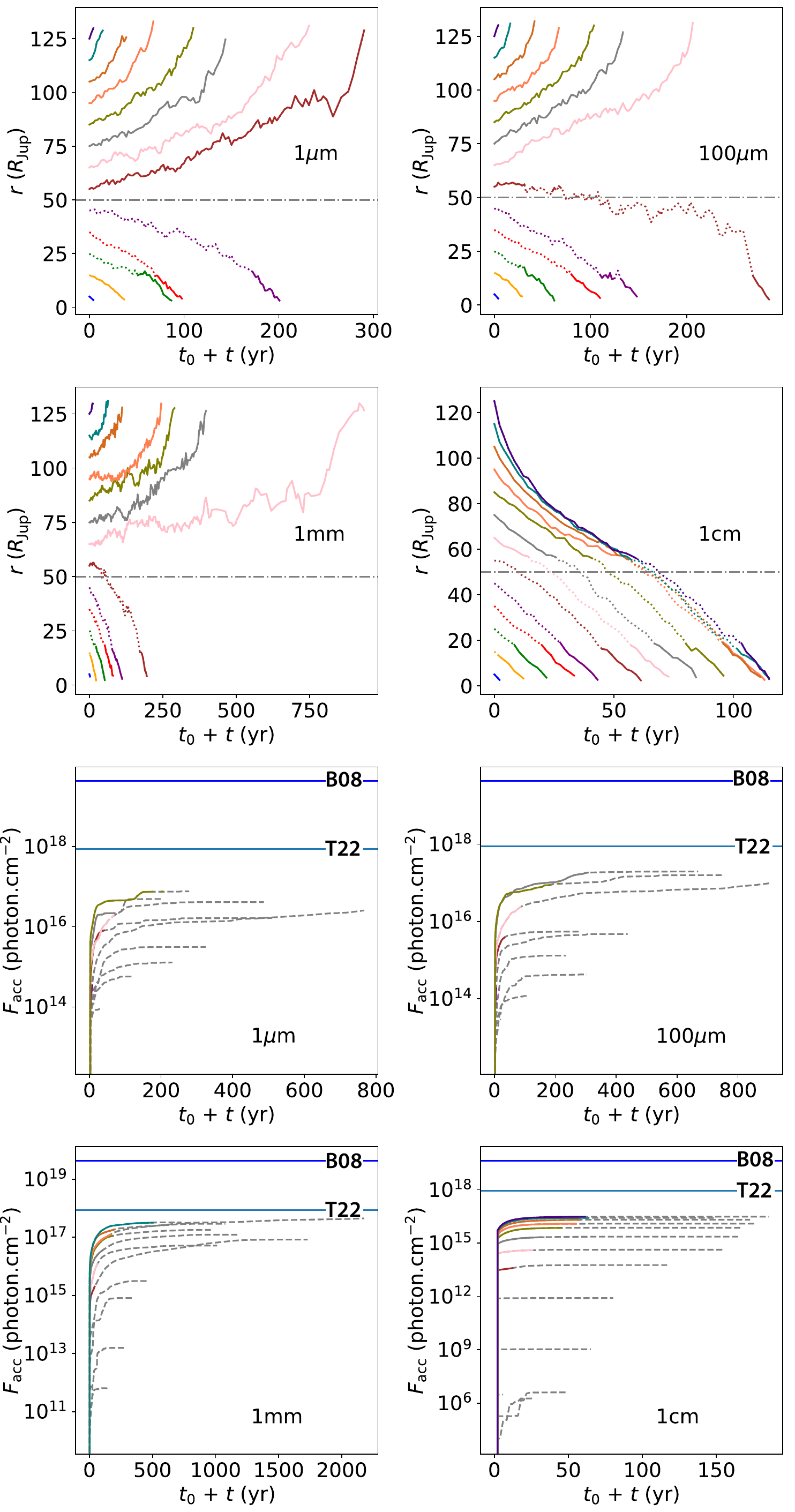}
\caption{{\it Top four panels}: median radial trajectories of 1 $\mu$m, 100 $\mu$m, 1 mm, and 1 cm particles as a function of time in our nominal CPD model. The particles are released one scale height above the CPD midplane at $t_{\rm 0}$ = 50 kyr. Median trajectories are shown at 10 $R_{\rm jup}$ intervals in the midplane, spanning from 5 to 135 $R_{\rm Jup}$ in the CPD. Dotted lines highlight portions of these trajectories that intersect the COM formation zone via thermal processing in the CPD. The horizontal dotted-dashed line indicates the location of $R_c$. {\it Bottom four panels}: average irradiation experienced by particles migrating inward within the CPD, shown as a function of time in the nominal CPD model. The two horizontal lines represent the irradiation thresholds discussed in the text, while the dashed curves indicate the fraction of trajectories where CH$_3$OH is in the vapor phase. T22 and B08 refer to the irradiation thresholds experimentally derived by \cite{tenelanda2022} and \cite{bossa2008}, respectively.}
\label{fig:50kyr}
\end{center}
\end{figure*}

Figure \ref{fig:50kyr} illustrates the time evolution of the median radial trajectories of particles of 1 $\mu$m, 100 $\mu$m, 1 mm and 1 cm released at $t_{\rm 0}$ = 50 kyr from the 14 different locations in the case of the nominal CPD model. As expected, most particles in the 1$\mu$m to 1mm size range are strongly coupled to the gas. Those released within the centrifugal radius $R_c$ (50~$R_{\rm Jup}$), follow the inward motion of the gas and reach Jupiter in less than 300 years. Conversely, most of the particles originating beyond $R_c$ are carried outward by the gas flow, reaching the outer edge of the CPD over a similar timescale. In contrast, 1cm particles experience significant gas drag and migrate inward more rapidly, reaching Jupiter in just over 100 yr. The particle trajectories also include segments that traverse the thermal processing region of NH$_3$-CO$_2$ ices, leading to the formation of COMs. This region lies between $\sim$20.6~$R_{\rm Jup}$ and R$_c$ within $\sim$300 years after $t_0$. If the Galilean moon embryos are still growing in this region after $t_0$~=~50 kyr into the CPD's evolution, they would be able to accrete COM--rich particles.

Figure~\ref{fig:50kyr} also presents the average irradiation experienced by particles along their radial trajectories for each batch of particles over time. Here, we consider only the particles migrating inward within the CPD, as these are more likely to be accreted by the forming Galilean moons. Two irradiation thresholds are indicated: a lower fluence of $8.64\times10^{17}$ photons~cm$^{-2}$ inferred from the laboratory experiments of \citet{tenelanda2022}, corresponding to the maximum UV dose explored in that study, and a higher fluence of $4.32\times10^{19}$ photons~cm$^{-2}$ reported by \citet{bossa2008}. Assuming the particles are primarily composed of CH$_3$OH ice, none reach the lower threshold, as they sublimate within a few hundred years at temperatures near 105 K \citep{Mousis2009}, early during their migration through the CPD. If, instead, the particles are composed of an NH$_3$:CO$_2$ ice mixture, the figure shows that thermal processing would convert the ice into COMs well before the irradiation threshold is reached. These results suggest that, under these compositional assumptions, thermal processing is the dominant pathway for COM formation in the CPD.


\begin{figure*}[!ht]
\begin{center}
\includegraphics[angle=0,width=10cm]{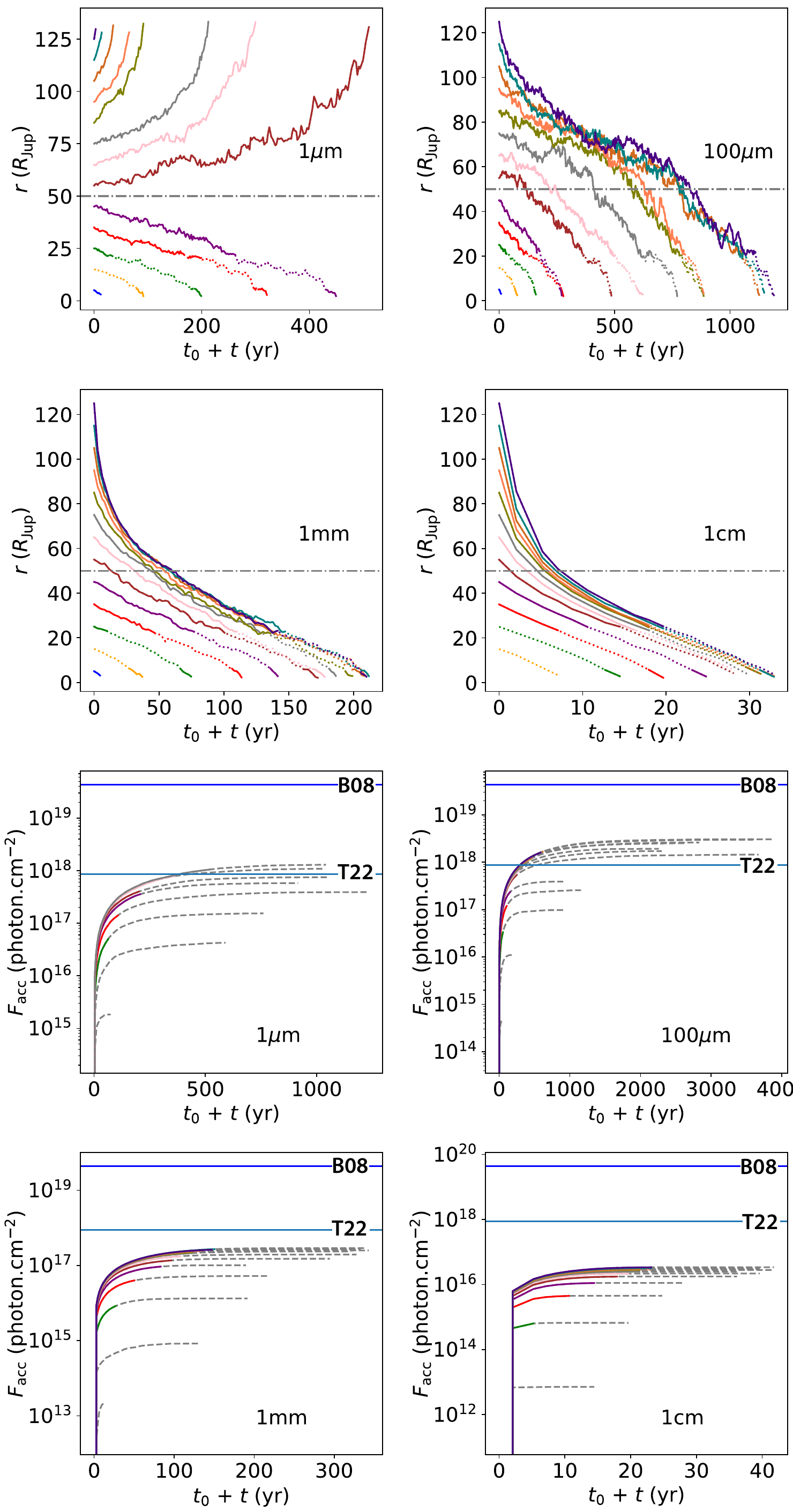}
\caption{Same as Fig. \ref{fig:50kyr}, but with particles released at $t_{\rm 0}$ = 100 kyr.}
\label{fig:100kyr}
\end{center}
\end{figure*}

\begin{figure*}[!ht]
\begin{center}
\includegraphics[angle=0,width=10cm]{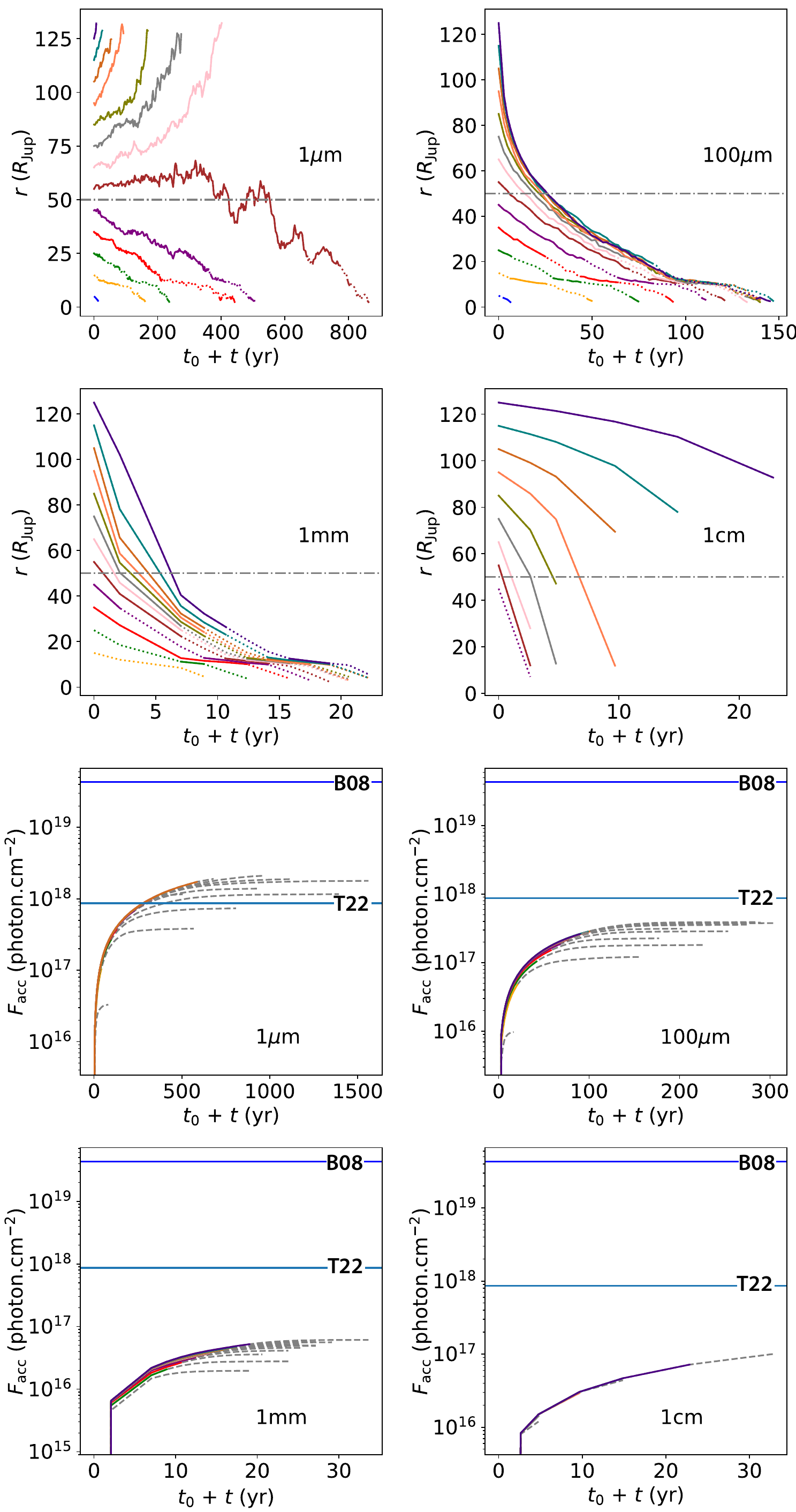}
\caption{Same as Fig. \ref{fig:50kyr}, but with particles released at $t_{\rm 0}$ = 150 kyr.}
\label{fig:150kyr}
\end{center}
\end{figure*}

The parameters adopted in Figs. \ref{fig:100kyr} and \ref{fig:150kyr} are identical to those used in Fig. \ref{fig:50kyr}, except that particles are released at $t_0 = 100$ kyr and $t_0 = 150$ kyr, respectively, during the evolution of the CPD. Since these release epochs occur later, the gas density in the CPD has decreased compared to the conditions at $t_0 = 50$ kyr.  As the gas density decreases, the Stokes number of the particles increases, since it is inversely proportional to the local gas density ($\mathrm{St} \propto 1/\rho_{\mathrm{gas}}$). This increase in the Stokes number results in weaker coupling between the particles and the gas. Consequently, at $t_0 = 100\,\mathrm{kyr}$, particles with sizes exceeding $100\,\mu\mathrm{m}$ are predominantly subject to gas drag, whereas at $t_0 = 150\,\mathrm{kyr}$, even particles as small as $1\,\mu\mathrm{m}$ exhibit inward drift, experiencing gas drag effects beyond $R_c$. At $t_0$ = 100 $\mathrm{kyr}$ and 150 $\mathrm{kyr}$, the thermal processing region shifts closer to Jupiter within the CPD, lying within approximately $\sim$$20\,R_{\mathrm{Jup}}$ during these stages of the disk's evolution. Ices composed of NH$_3$:CO$_2$, whether carried by grains influenced by gas drag or by the smallest particles tightly coupled to the gas and released below $R_c$, transit through this region. During this passage, thermal processing may convert these ices into COMs.

Interestingly, Fig.~\ref{fig:100kyr} indicates that particles released at $t_0 = 100\,\mathrm{kyr}$ beyond approximately 75 and 125 $R_{\mathrm{Jup}}$, with sizes of 1 $\mu\mathrm{m}$ and 100 $\mu\mathrm{m}$, respectively, follow trajectories that exceed the irradiation threshold defined by \citet{tenelanda2022}. However, as the figure shows, most of these particles lose their CH$_3$OH ice due to sublimation before reaching this threshold. A similar trend is observed in Fig.~\ref{fig:150kyr}, where the computed trajectories of 1--$\mu\mathrm{m}$ particles released at $t_0 = 150\,\mathrm{kyr}$ beyond approximately 85 $R_{\mathrm{Jup}}$ in the CPD reach the irradiation threshold, while most of them have lost their CH$_3$OH ice by that time.

\begin{figure*}
\centering
\includegraphics[width=15cm]{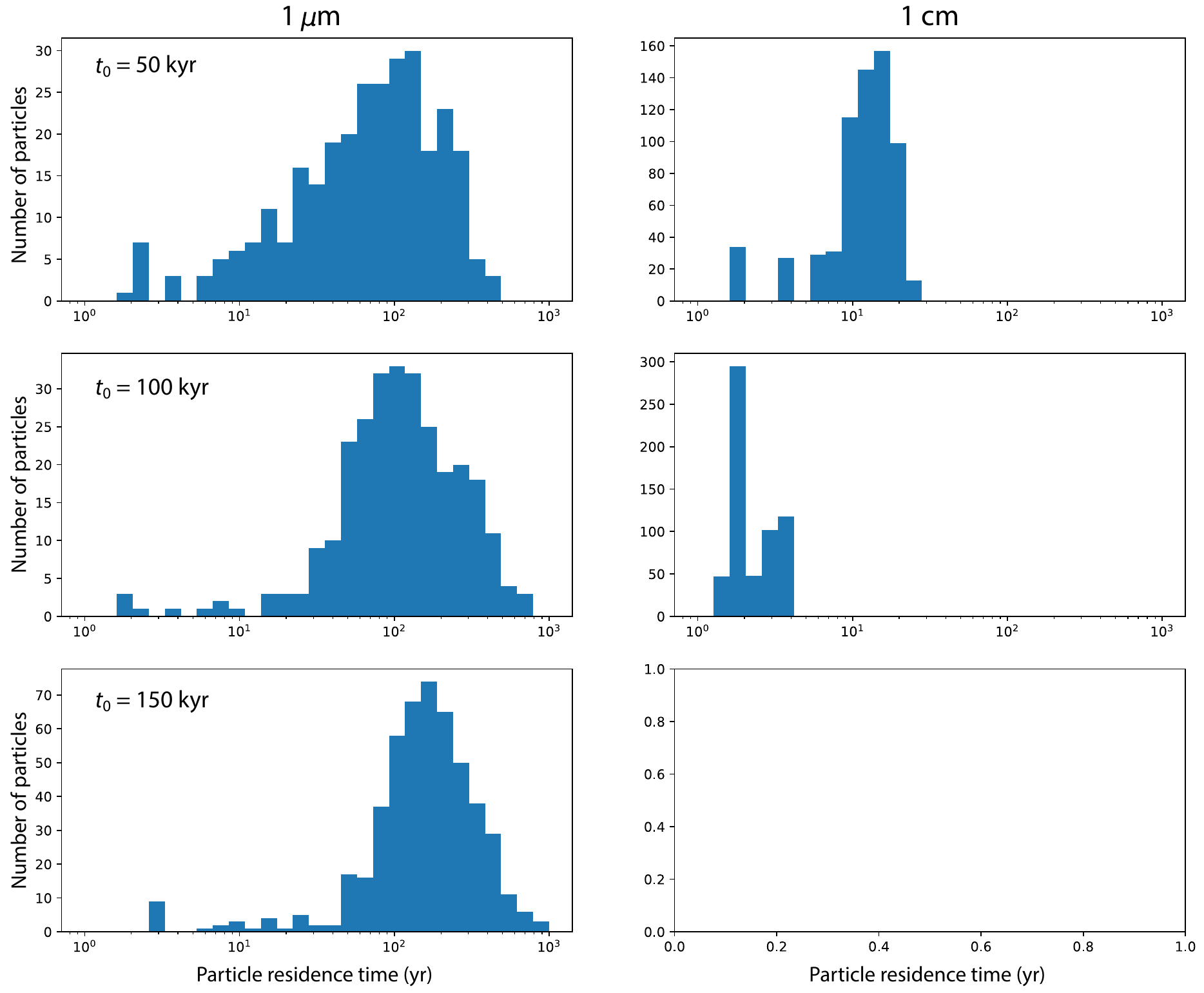}
\caption{Time evolution of the number of 1 $\mu$m and 1 cm particles remaining in the COM stability region ($\sim$80--260 K), relative to an initial set of 800 particles released in our nominal CPD model. From top to bottom, the particles were initially uniformly distributed between 5 and 135 $R_{\rm jup}$ at one scale height above the CPD midplane, with starting epochs $t_{0} = 50$, 100, and 150~kyr, respectively. When $t_{0} \simeq 150$~kyr, 1~cm particles migrate on timescales shorter than the adopted CPD evolution timestep (1~yr).}
\label{fig:distrib}
\end{figure*}

Figure~\ref{fig:distrib} shows the distribution of residence times for particles within the COM thermal stability region, considering particles of sizes 1~$\mu$m and 1~cm, released into the CPD at $t_0$ = 50, 100, and 150~kyr after its formation. Both particle sizes have short residence times, with the largest particles being removed fastest. For 1~$\mu$m particles, the residence time distribution peaks at around 100~yr at $t_0$ = 50~kyr and shifts to approximately 200~yr at $t_0$ = 100 and 150~kyr. In contrast, 1~cm particles exhibit much shorter residence times. Those released at $t_0$ = 50~kyr typically remain within the thermal stability region of COMs for less than 30~yr, decreasing to under 4~yr at $t_0$ = 100~kyr. None of the 1~cm particles released at $t_0$ = 150~kyr remain in the stability region for more than 1~yr, which corresponds to the temporal resolution of the iterative scheme of our model.

\section{Discussion} 
\label{sec:sec4}

Overall, our simulations suggest that particles of varying sizes, released at different points during the evolution of the CPD, pass through regions where temperatures are high enough to thermally process ices, particularly NH$_3$:CO$_2$ ices, into complex COMs. This transformation occurs before significant irradiation, indicating that thermal processes, rather than photochemical reactions, primarily drive the formation of COMs in this environment. Furthermore, particles typically reach the methanol iceline before they can absorb a sufficient number of photons, making the formation of COMs through irradiation a challenging and intricate process in CPDs.

The results of our study have been tested across a large range of parameter sets. For instance, an increase in particle density leads to a faster inward drift. Specifically, all 1 cm particles with a density of 3 g.cm$^{-3}$ reach Jupiter in less than 40 yr, compared to $\sim$100 yr for particles with a density of 1 g.cm$^{-3}$, assuming $t_0$ = 50 kyr. This indicates that the formation of COMs via irradiation is more challenging, as these particles receive lower irradiation doses. Nevertheless, they still traverse the thermal processing zone. Reducing the value of $R_c$, for instance to 17~$R_{\mathrm{Jup}}$, has little effect on the irradiation doses received by the smallest particles. However, it reduces the efficiency of COM formation via thermal processing. As $R_c$ moves closer to Jupiter, fewer small particles are entrained in the inward gas flow and transported through the thermal processing region. In contrast, the dynamics of larger particles, which are primarily governed by gas drag, remain largely unaffected by variations in $R_c$ in terms of both irradiation and thermal processing.

An increase in the turbulent viscosity parameter, $\alpha$, of the CPD results in a lower surface density and a lower midplane temperature. This limits the spatial extent of regions where thermal processing of particles can occur at a given stage of the CPD's evolution. Furthermore, increasing $\alpha$ reduces the likelihood of COM formation by irradiation. For example, when $\alpha = 10^{-2}$, COMs are not produced by irradiation because the lower gas density increases the Stokes number of small particles, which weakens their coupling to the gas. Consequently, these particles drift inward more rapidly toward Jupiter, shortening their residence time in the outer CPD where irradiation is effective. Decreasing the CPD accretion rate has a similar effect to increasing the turbulent viscosity parameter, $\alpha$, resulting in lower surface densities and midplane temperatures. Decreasing the CPD's accretion rate by an order of magnitude, for instance, reduces the particles' residence time within the disk by about half, thereby limiting their exposure to conditions conducive to COM formation via irradiation.

The limited availability of laboratory photochemical data introduces unavoidable uncertainties in our modeling. Our approach relies on the most recent UV-irradiation experiments, which probe a relatively narrow wavelength range and may therefore bias the predicted reaction pathways and yields of COMs. Consistent with previous studies adopting similar formalisms to evaluate irradiation in PPDs \citep{Mu02,Ob09,Ben24,Benest_Couzinou_2025}, our chemical model includes only a restricted set of formation and destruction processes and neglects hydrogenation reactions \citep{Lin11}, as well as secondary UV photons generated by cosmic-ray interactions with H$_2$ gas \citep{Pr83}. Reactions involving additional molecular species and the implementation of more comprehensive photochemical networks \citep{Tak22,Ochiai24} are also not considered, which may limit the model’s ability to fully capture the chemical evolution within the CPD.

It is worth noting that, thermal and photochemical processes operate through fundamentally different mechanisms: thermal reactions primarily drive molecular reorganization in warm regions, whereas photochemistry dominates in UV-irradiated layers. Future laboratory studies spanning a broader range of photon energies and irradiation conditions will be essential to further refine chemical models and improve quantitative predictions. Nevertheless, we expect our main qualitative conclusions to remain robust, as they are governed by the distinct roles of thermal and photochemical processing rather than by the precise efficiencies of individual reaction pathways.

The present study intentionally isolates the chemical evolution of COMs, whereas in astrophysical environments these species coexist with water ice, silicates, and carbonaceous grains that can significantly influence their stability and reactivity. When CH$_3$OH is embedded in an H$_2$O-rich matrix, UV irradiation of the ice substantially modifies both radical production and subsequent chemical evolution \citep{Ob10}. Increasing the H$_2$O fraction enhances the photodestruction efficiency of volatile constituents and shifts the product distribution toward more oxygen-rich species while simultaneously promoting radical trapping in the ice matrix and disproportionately inhibiting OH diffusion relative to CH$_3$ and HCO. These competing effects imply that, although H$_2$O-rich ices can efficiently generate radicals, the formation pathways and yields of complex organic molecules are strongly regulated by mixture-dependent radical mobility, emphasizing the need to account for H$_2$O-controlled diffusion and trapping when interpreting or modeling COM formation in astrophysical ices.

Adsorption onto mineral and carbonaceous surfaces can also modify reaction barriers and promote catalytic organic processing \citep{He09,Mi16,Cu17}, while incorporation within mixed ice–dust aggregates can reduce the effective UV flux reaching embedded molecules through geometric and optical shielding, thereby limiting UV photodissociation \citep{Po20} and affect their retention and release during thermal desorption and sublimation \citep{Fa11}. Future work will explore the inclusion of simplified adsorption–desorption processes in the photochemical–thermal framework to examine their potential influence on the survival and distribution of COMs in the Jovian CPD.

Although the total irradiation dose was compared with laboratory conditions, the significantly lower irradiation rate in the CPD likely limits molecular complexity, as radicals formed under such conditions have more time to recombine into their precursors. Second, the treatment of opacity and UV irradiation introduces additional limitations. In particular, the Rosseland mean opacity adopted in the CPD model, being frequency-averaged, does not adequately represent UV--specific processes, as already noted in the context of PPDs \citep{Ben24}. Although this limitation could lead to an underestimation of UV penetration and thus of irradiation-driven chemistry, it is partially compensated by the assumption that the adopted interstellar UV flux, $F_0$, represents a lower bound. Higher irradiation levels are plausible during the early evolution of the Solar System, especially during the CPD phase. Such elevated fluxes could significantly enhance the formation of COMs via irradiation-driven processes. A likely scenario for achieving high UV fluxes involves the formation of the Sun within a dense stellar cluster, in close proximity ($\le$0.03--0.05~pc) to one or more O-type stars. These stars emit intense FUV radiation, capable of generating flux levels as high as $G_0$ = 30{,}000 \citep{Adams04, Fat08}. Additionally, meteoritic evidence, including the presence of short-lived radionuclides such as $^{26}$Al and $^{60}$Fe, suggests that the early solar system experienced significant external irradiation, possibly from nearby supernovae \citep{Lich16}. These extreme, yet observationally supported conditions, are consistent with a clustered origin for the Sun.

Our calculations suggest that, if COM--rich particles were present in the CPD, either formed in situ or delivered from the PPD, the Galilean moons could have incorporated these materials during their accretion. However, accounting for their present-day physical states implies that each moon must have formed under distinct conditions, leading to varying efficiencies in retaining solid-phase COMs within their interiors. The thermal environment during accretion was likely influenced by multiple factors, including the sizes of the moons and their impactors, the local disk temperature, and the duration of the accretion process \citep{canup2002,Bierson_Nimmo_2020,Bennacer_2025}. 

The combination of accretional heating and high ambient temperatures in the inner CPD may have led to high-temperature accretion of Io and Europa \citep{Bierson_Nimmo_2020}, causing the destruction of COMs. However, more recent studies suggest that Europa may have formed under cooler conditions \citep{Trinh_2023, Petricca_2025}, where the gravitational energy delivered by impactors was minor compared to the background thermal input from the CPD, due to a prolonged accretion period \citep{canup2002}. In this context, assuming a low heat dissipation efficiency ($h$ $\sim$0.01), the Europa accretion temperature would closely follow the local CPD temperature, estimated between 200 and 300 K at its formation site. Thus, COMs could have remained stable within Europa if it accreted slowly, typically over several million years, and at a distance greater from Jupiter than its current orbit.

Both Ganymede and Callisto are thought to have formed beyond the snow line in the colder regions of the CPD. Depending on the size distribution of the impactors and the duration of accretion, Ganymede’s accretional history could have been warm or cold \citep{Barr_Canup_2008, Bierson_Nimmo_2020, Bjonnes22, Bennacer_2025}. \citet{Bennacer_2025} recently showed that Ganymede could avoid global melting only if it accreted slowly and primarily from small particles with radii $r_{\text{imp}} \lesssim 100$ m. Even in scenarios involving a substantial fraction of kilometer-sized impactors, thermal models indicate that elevated temperatures would largely be confined to the final stages of accretion. This suggests that Ganymede may have retained a significant fraction of its primordial COMs. Callisto, by contrast, is widely believed to have formed under cold accretion conditions, consistent with its partially differentiated interior and incomplete ice-rock segregation \citep{Anderson_2001, Barr_Canup_2008, Bennacer_2025}. Consequently, both moons may have retained substantial fractions of the COMs acquired during their formation.


\bibliography{CPD}

@ARTICLE{Mu02,
       author = {{Mu{\~n}oz Caro}, G.~M. and {Meierhenrich}, U.~J. and {Schutte}, W.~A. and {Barbier}, B. and {Arcones Segovia}, A. and {Rosenbauer}, H. and {Thiemann}, W.~H.-P. and {Brack}, A. and {Greenberg}, J.~M.},
        title = "{Amino acids from ultraviolet irradiation of interstellar ice analogues}",
      journal = {\nat},
         year = 2002,
        month = mar,
       volume = {416},
       number = {6879},
        pages = {403-406},
          doi = {10.1038/416403a},
       adsurl = {https://ui.adsabs.harvard.edu/abs/2002Natur.416..403M}
}

@ARTICLE{Ob10,
       author = {{{\"O}berg}, Karin I. and {van Dishoeck}, Ewine F. and {Linnartz}, Harold and {Andersson}, Stefan},
        title = "{The Effect of H$_{2}$O on Ice Photochemistry}",
      journal = {\apj},
         year = 2010,
        month = aug,
       volume = {718},
       number = {2},
        pages = {832-840},
          doi = {10.1088/0004-637X/718/2/832},
archivePrefix = {arXiv},
       eprint = {1006.2190},
 primaryClass = {astro-ph.GA},
       adsurl = {https://ui.adsabs.harvard.edu/abs/2010ApJ...718..832O}
}

@ARTICLE{Mi16,
       author = {{Minissale}, M. and {Dulieu}, F. and {Cazaux}, S. and {Hocuk}, S.},
        title = "{Dust as interstellar catalyst. I. Quantifying the chemical desorption process}",
      journal = {\aap},
         year = 2016,
        month = jan,
       volume = {585},
          eid = {A24},
        pages = {A24},
          doi = {10.1051/0004-6361/201525981},
archivePrefix = {arXiv},
       eprint = {1510.03218},
 primaryClass = {astro-ph.SR},
       adsurl = {https://ui.adsabs.harvard.edu/abs/2016A&A...585A..24M}
}

@ARTICLE{Cu17,
       author = {{Cuppen}, H.~M. and {Walsh}, C. and {Lamberts}, T. and {Semenov}, D. and {Garrod}, R.~T. and {Penteado}, E.~M. and {Ioppolo}, S.},
        title = "{Grain Surface Models and Data for Astrochemistry}",
      journal = {\ssr},
         year = 2017,
        month = oct,
       volume = {212},
       number = {1-2},
        pages = {1-58},
          doi = {10.1007/s11214-016-0319-3},
       adsurl = {https://ui.adsabs.harvard.edu/abs/2017SSRv..212....1C}
}

@ARTICLE{He09,
       author = {{Herbst}, Eric and {van Dishoeck}, Ewine F.},
        title = "{Complex Organic Interstellar Molecules}",
      journal = {\araa},
         year = 2009,
        month = sep,
       volume = {47},
       number = {1},
        pages = {427-480},
          doi = {10.1146/annurev-astro-082708-101654},
       adsurl = {https://ui.adsabs.harvard.edu/abs/2009ARA&A..47..427H}
}

@ARTICLE{Po20,
       author = {{Potapov}, Alexey and {J{\"a}ger}, Cornelia and {Henning}, Thomas},
        title = "{Ice Coverage of Dust Grains in Cold Astrophysical Environments}",
      journal = {\prl},
         year = 2020,
        month = jun,
       volume = {124},
       number = {22},
          eid = {221103},
        pages = {221103},
          doi = {10.1103/PhysRevLett.124.221103},
archivePrefix = {arXiv},
       eprint = {2005.00757},
 primaryClass = {astro-ph.GA},
       adsurl = {https://ui.adsabs.harvard.edu/abs/2020PhRvL.124v1103P}
}

@ARTICLE{
Fa11,
       author = {{Fayolle}, E.~C. and {{\"O}berg}, K.~I. and {Cuppen}, H.~M. and {Visser}, R. and {Linnartz}, H.},
        title = "{Laboratory H$_{2}$O:CO$_{2}$ ice desorption data: entrapment dependencies and its parameterization with an extended three-phase model}",
      journal = {\aap},
         year = 2011,
        month = may,
       volume = {529},
          eid = {A74},
        pages = {A74},
          doi = {10.1051/0004-6361/201016121},
archivePrefix = {arXiv},
       eprint = {1108.6055},
 primaryClass = {astro-ph.EP},
       adsurl = {https://ui.adsabs.harvard.edu/abs/2011A&A...529A..74F}
}

@ARTICLE{Ob09,
       author = {{{\"O}berg}, K.~I. and {Garrod}, R.~T. and {van Dishoeck}, E.~F. and {Linnartz}, H.},
        title = "{Formation rates of complex organics in UV irradiated CH\_3OH-rich ices. I. Experiments}",
      journal = {\aap},
         year = 2009,
        month = sep,
       volume = {504},
       number = {3},
        pages = {891-913},
          doi = {10.1051/0004-6361/200912559},
archivePrefix = {arXiv},
       eprint = {0908.1169},
 primaryClass = {astro-ph.GA},
       adsurl = {https://ui.adsabs.harvard.edu/abs/2009A&A...504..891O}
}

@ARTICLE{Ben25,
       author = {{Couzinou}, Tom Benest and {Moulanier}, Aliz{\'e}e Amsler and {Mousis}, Olivier},
        title = "{Delivery of complex organic molecules to the system of Jupiter}",
      journal = {\mnras},
         year = 2025,
        month = nov,
          doi = {10.1093/mnras/staf2074},
       adsurl = {https://ui.adsabs.harvard.edu/abs/2025MNRAS.tmp.1992C}
}

@ARTICLE{Wa14,
       author = {{Walsh}, Catherine and {Millar}, Tom. J. and {Nomura}, Hideko and {Herbst}, Eric and {Widicus Weaver}, Susanna and {Aikawa}, Yuri and {Laas}, Jacob C. and {Vasyunin}, Anton I.},
        title = "{Complex organic molecules in protoplanetary disks}",
      journal = {\aap},
         year = 2014,
        month = mar,
       volume = {563},
          eid = {A33},
        pages = {A33},
          doi = {10.1051/0004-6361/201322446},
archivePrefix = {arXiv},
       eprint = {1403.0390},
 primaryClass = {astro-ph.EP},
       adsurl = {https://ui.adsabs.harvard.edu/abs/2014A&A...563A..33W}
}

@ARTICLE{Po24,
       author = {{Poulet}, F. and {Piccioni}, G. and {Langevin}, Y. and {Dumesnil}, C. and {Tommasi}, L. and {Carlier}, V. and {Filacchione}, G. and {Amoroso}, M. and {Arondel}, A. and {D'Aversa}, E. and {Barbis}, A. and {Bini}, A. and {Bols{\'e}e}, D. and {Bousquet}, P. and {Caprini}, C. and {Carter}, J. and {Dubois}, J. -P. and {Condamin}, M. and {Couturier}, S. and {Dassas}, K. and {Dexet}, M. and {Fletcher}, L. and {Grassi}, D. and {Guerri}, I. and {Haffoud}, P. and {Larigauderie}, C. and {Le Du}, M. and {Mugnuolo}, R. and {Pilato}, G. and {Rossi}, M. and {Stefani}, S. and {Tosi}, F. and {Vincendon}, M. and {Zambelli}, M. and {Arnold}, G. and {Bibring}, J. -P. and {Biondi}, D. and {Boccaccini}, A. and {Brunetto}, R. and {Carapelle}, A. and {Cisneros Gonz{\'a}lez}, M. and {Hannou}, C. and {Karatekin}, O. and {Le Cle'ch}, J. -C. and {Leyrat}, C. and {Migliorini}, A. and {Nathues}, A. and {Rodriguez}, S. and {Saggin}, B. and {Sanchez-Lavega}, A. and {Schmitt}, B. and {Seignovert}, B. and {Sordini}, R. and {Stephan}, K. and {Tobie}, G. and {Zambon}, F. and {Adriani}, A. and {Altieri}, F. and {Bockel{\'e}e}, D. and {Capaccioni}, F. and {De Angelis}, S. and {De Sanctis}, M. -C. and {Drossart}, P. and {Fouchet}, T. and {G{\'e}rard}, J. -C. and {Grodent}, D. and {Ignatiev}, N. and {Irwin}, P. and {Ligier}, N. and {Manaud}, N. and {Mangold}, N. and {Mura}, A. and {Pilorget}, C. and {Quirico}, E. and {Renotte}, E. and {Strazzulla}, G. and {Turrini}, D. and {Vandaele}, A. -C. and {Carli}, C. and {Ciarniello}, M. and {Guerlet}, S. and {Lellouch}, E. and {Mancarella}, F. and {Morbidelli}, A. and {Le Mou{\'e}lic}, S. and {Raponi}, A. and {Sindoni}, G. and {Snels}, M.},
        title = "{Moons and Jupiter Imaging Spectrometer (MAJIS) on Jupiter Icy Moons Explorer (JUICE)}",
      journal = {\ssr},
         year = 2024,
        month = mar,
       volume = {220},
       number = {3},
          eid = {27},
        pages = {27},
          doi = {10.1007/s11214-024-01057-2},
       adsurl = {https://ui.adsabs.harvard.edu/abs/2024SSRv..220...27P}
}

@INPROCEEDINGS{Ha23,
       author = {{Hartogh}, Paul},
        title = "{The Submillimeter Wave Instrument (SWI) on the JUpiter ICy moons Explorer (JUICE)}",
    booktitle = {AAS/Division for Planetary Sciences Meeting Abstracts \#55},
         year = 2023,
       series = {AAS/Division for Planetary Sciences Meeting Abstracts},
       volume = {55},
        month = oct,
          eid = {201.02},
        pages = {201.02},
       adsurl = {https://ui.adsabs.harvard.edu/abs/2023DPS....5520102H}
}

@ARTICLE{Wa24,
       author = {{Waite}, J.~H. and {Burch}, J.~L. and {Brockwell}, T.~G. and {Young}, D.~T. and {Miller}, G.~P. and {Persyn}, S.~C. and {Stone}, J.~M. and {Wilson}, P. and {Miller}, K.~E. and {Glein}, C.~R. and {Perryman}, R.~S. and {McGrath}, M.~A. and {Bolton}, S.~J. and {McKinnon}, W.~B. and {Mousis}, O. and {Sephton}, M.~A. and {Shock}, E.~L. and {Choukroun}, M. and {Teolis}, B.~D. and {Wyrick}, D.~Y. and {Zolotov}, M.~Y. and {Ray}, C. and {Magoncelli}, A.~L. and {Raffanti}, R.~R. and {Thorpe}, R.~L. and {Bouquet}, A. and {Salter}, T.~L. and {Robinson}, K.~J. and {Urdiales}, C. and {Tyler}, Y.~D. and {Dirks}, G.~J. and {Beebe}, C.~R. and {Fugett}, D.~A. and {Alexander}, J.~A. and {Hanley}, J.~J. and {Moorhead-Rosenberg}, Z.~A. and {Franke}, K.~A. and {Pickens}, K.~S. and {Focia}, R.~J. and {Magee}, B.~A. and {Hoeper}, P.~J. and {Aaron}, D.~P. and {Thompson}, S.~L. and {Persson}, K.~B. and {Blase}, R.~C. and {Dunn}, G.~F. and {Killough}, R.~L. and {De Los Santos}, A. and {Rickerson}, R.~J. and {Siegmund}, O.~H.~W.},
        title = "{MASPEX-Europa: The Europa Clipper Neutral Gas Mass Spectrometer Investigation}",
      journal = {\ssr},
         year = 2024,
        month = apr,
       volume = {220},
       number = {3},
          eid = {30},
        pages = {30},
          doi = {10.1007/s11214-024-01061-6},
       adsurl = {https://ui.adsabs.harvard.edu/abs/2024SSRv..220...30W}
}

@INPROCEEDINGS{Wu18,
       author = {{Wurz}, Peter and {Meyer}, Stefan and {Galli}, Andr{\'e} and {Tulej}, Marek and {Vorburger}, Audrey and {Lasi}, Davide and {Piazza}, Daniele and {L{\"u}thi}, Matthias and {Brandt}, Pontus and {Barabash}, Stas},
        title = "{The Neutral Gas and Ion Mass Spectrometer of the PEP Experiment on the JUICE Mission}",
    booktitle = {EGU General Assembly Conference Abstracts},
         year = 2018,
       series = {EGU General Assembly Conference Abstracts},
        month = apr,
        pages = {10091},
       adsurl = {https://ui.adsabs.harvard.edu/abs/2018EGUGA..2010091W}
}

@ARTICLE{Bec24,
       author = {{Becker}, T.~M. and {Zolotov}, M.~Y. and {Gudipati}, M.~S. and {Soderblom}, J.~M. and {McGrath}, M.~A. and {Henderson}, B.~L. and {Hedman}, M.~M. and {Choukroun}, M. and {Clark}, R.~N. and {Chivers}, C. and {Wolfenbarger}, N.~S. and {Glein}, C.~R. and {Castillo-Rogez}, J.~C. and {Mousis}, O. and {Scanlan}, K.~M. and {Diniega}, S. and {Seelos}, F.~P. and {Goode}, W. and {Postberg}, F. and {Grima}, C. and {Hsu}, H. -W. and {Roth}, L. and {Trumbo}, S.~K. and {Miller}, K.~E. and {Chan}, K. and {Paranicas}, C. and {Brooks}, S.~M. and {Soderlund}, K.~M. and {McKinnon}, W.~B. and {Hibbitts}, C.~A. and {Smith}, H.~T. and {Molyneux}, P.~M. and {Gladstone}, G.~R. and {Cable}, M.~L. and {Ulibarri}, Z.~E. and {Teolis}, B.~D. and {Horanyi}, M. and {Jia}, X. and {Leonard}, E.~J. and {Hand}, K.~P. and {Vance}, S.~D. and {Howell}, S.~M. and {Quick}, L.~C. and {Mishra}, I. and {Rymer}, A.~M. and {Briois}, C. and {Blaney}, D.~L. and {Raut}, U. and {Waite}, J.~H. and {Retherford}, K.~D. and {Shock}, E. and {Withers}, P. and {Westlake}, J.~H. and {Jun}, I. and {Mandt}, K.~E. and {Buratti}, B.~J. and {Korth}, H. and {Pappalardo}, R.~T. and {Europa Clipper Composition Working Group}},
        title = "{Exploring the Composition of Europa with the Upcoming Europa Clipper Mission}",
      journal = {\ssr},
         year = 2024,
        month = aug,
       volume = {220},
       number = {5},
          eid = {49},
        pages = {49},
          doi = {10.1007/s11214-024-01069-y},
       adsurl = {https://ui.adsabs.harvard.edu/abs/2024SSRv..220...49B}
}

@ARTICLE{papp24,
       author = {{Pappalardo}, Robert T. and {Buratti}, Bonnie J. and {Korth}, Haje and {Senske}, David A. and {Blaney}, Diana L. and {Blankenship}, Donald D. and {Burch}, James L. and {Christensen}, Philip R. and {Kempf}, Sascha and {Kivelson}, Margaret G. and {Mazarico}, Erwan and {Retherford}, Kurt D. and {Turtle}, Elizabeth P. and {Westlake}, Joseph H. and {Paczkowski}, Brian G. and {Ray}, Trina L. and {Kampmeier}, Jennifer and {Craft}, Kate L. and {Howell}, Samuel M. and {Klima}, Rachel L. and {Leonard}, Erin J. and {Matiella Novak}, Alexandra and {Phillips}, Cynthia B. and {Daubar}, Ingrid J. and {Blacksberg}, Jordana and {Brooks}, Shawn M. and {Choukroun}, Mathieu N. and {Cochrane}, Corey J. and {Diniega}, Serina and {Elder}, Catherine M. and {Ernst}, Carolyn M. and {Gudipati}, Murthy S. and {Luspay-Kuti}, Adrienn and {Piqueux}, Sylvain and {Rymer}, Abigail M. and {Roberts}, James H. and {Steinbr{\"u}gge}, Gregor and {Cable}, Morgan L. and {Scully}, Jennifer E.~C. and {Castillo-Rogez}, Julie C. and {Hay}, Hamish C.~F.~C. and {Persaud}, Divya M. and {Glein}, Christopher R. and {McKinnon}, William B. and {Moore}, Jeffrey M. and {Raymond}, Carol A. and {Schroeder}, Dustin M. and {Vance}, Steven D. and {Wyrick}, Danielle Y. and {Zolotov}, Mikhail Y. and {Hand}, Kevin P. and {Nimmo}, Francis and {McGrath}, Melissa A. and {Spencer}, John R. and {Lunine}, Jonathan I. and {Paty}, Carol S. and {Soderblom}, Jason M. and {Collins}, Geoffrey C. and {Schmidt}, Britney E. and {Rathbun}, Julie A. and {Shock}, Everett L. and {Becker}, Tracy C. and {Hayes}, Alexander G. and {Prockter}, Louise M. and {Weiss}, Benjamin P. and {Hibbitts}, Charles A. and {Moussessian}, Alina and {Brockwell}, Timothy G. and {Hsu}, Hsiang-Wen and {Jia}, Xianzhe and {Gladstone}, G. Randall and {McEwen}, Alfred S. and {Patterson}, G. Wesley and {McNutt}, Ralph L. and {Evans}, Jordan P. and {Larson}, Timothy W. and {Cangahuala}, L. Alberto and {Havens}, Glen G. and {Buffington}, Brent B. and {Bradley}, Ben and {Campagnola}, Stefano and {Hardman}, Sean H. and {Srinivasan}, Jeffrey M. and {Short}, Kendra L. and {Jedrey}, Thomas C. and {St. Vaughn}, Joshua A. and {Clark}, Kevin P. and {Vertesi}, Janet and {Niebur}, Curt},
        title = "{Science Overview of the Europa Clipper Mission}",
      journal = {\ssr},
         year = 2024,
        month = may,
       volume = {220},
       number = {4},
          eid = {40},
        pages = {40},
          doi = {10.1007/s11214-024-01070-5},
       adsurl = {https://ui.adsabs.harvard.edu/abs/2024SSRv..220...40P}
}

@INCOLLECTION{Ha09,
       author = {{Hand}, K.~P. and {Chyba}, C.~F. and {Priscu}, J.~C. and {Carlson}, R.~W. and {Nealson}, K.~H.},
        title = "{Astrobiology and the Potential for Life on Europa}",
    booktitle = {Europa},
         year = 2009,
       editor = {{Pappalardo}, Robert T. and {McKinnon}, William B. and {Khurana}, Krishan K.},
        pages = {589},
       adsurl = {https://ui.adsabs.harvard.edu/abs/2009euro.book..589H}
}

@ARTICLE{Mc08,
       author = {{McKay}, Christopher P. and {Porco Carolyn C.} and {Altheide}, Travis and {Davis}, Wanda L. and {Kral}, Timothy A.},
        title = "{The Possible Origin and Persistence of Life on Enceladus and Detection of Biomarkers in the Plume}",
      journal = {Astrobiology},
         year = 2008,
        month = oct,
       volume = {8},
       number = {5},
        pages = {909-919},
          doi = {10.1089/ast.2008.0265},
       adsurl = {https://ui.adsabs.harvard.edu/abs/2008AsBio...8..909M}
}

@ARTICLE{tocard13,
       author = {{Grasset}, O. and {Dougherty}, M.~K. and {Coustenis}, A. and {Bunce}, E.~J. and {Erd}, C. and {Titov}, D. and {Blanc}, M. and {Coates}, A. and {Drossart}, P. and {Fletcher}, L.~N. and {Hussmann}, H. and {Jaumann}, R. and {Krupp}, N. and {Lebreton}, J. -P. and {Prieto-Ballesteros}, O. and {Tortora}, P. and {Tosi}, F. and {Van Hoolst}, T.},
        title = "{JUpiter ICy moons Explorer (JUICE): An ESA mission to orbit Ganymede and to characterise the Jupiter system}",
      journal = {\planss},
         year = 2013,
        month = apr,
       volume = {78},
        pages = {1-21},
          doi = {10.1016/j.pss.2012.12.002},
       adsurl = {https://ui.adsabs.harvard.edu/abs/2013P&SS...78....1G}
}

@ARTICLE{Saur15,
       author = {{Saur}, Joachim and {Duling}, Stefan and {Roth}, Lorenz and {Jia}, Xianzhe and {Strobel}, Darrell F. and {Feldman}, Paul D. and {Christensen}, Ulrich R. and {Retherford}, Kurt D. and {McGrath}, Melissa A. and {Musacchio}, Fabrizio and {Wennmacher}, Alexandre and {Neubauer}, Fritz M. and {Simon}, Sven and {Hartkorn}, Oliver},
        title = "{The search for a subsurface ocean in Ganymede with Hubble Space Telescope observations of its auroral ovals}",
      journal = {Journal of Geophysical Research (Space Physics)},
         year = 2015,
        month = mar,
       volume = {120},
       number = {3},
        pages = {1715-1737},
          doi = {10.1002/2014JA020778},
       adsurl = {https://ui.adsabs.harvard.edu/abs/2015JGRA..120.1715S}
}

@ARTICLE{Vance23,
       author = {{Vance}, Steven D. and {Craft}, Kathleen L. and {Shock}, Everett and {Schmidt}, Britney E. and {Lunine}, Jonathan and {Hand}, Kevin P. and {McKinnon}, William B. and {Spiers}, Elizabeth M. and {Chivers}, Chase and {Lawrence}, Justin D. and {Wolfenbarger}, Natalie and {Leonard}, Erin J. and {Robinson}, Kirtland J. and {Styczinski}, Marshall J. and {Persaud}, Divya M. and {Steinbr{\"u}gge}, Gregor and {Zolotov}, Mikhail Y. and {Quick}, Lynnae C. and {Scully}, Jennifer E.~C. and {Becker}, Tracy M. and {Howell}, Samuel M. and {Clark}, Roger N. and {Dombard}, Andrew J. and {Glein}, Christopher R. and {Mousis}, Olivier and {Sephton}, Mark A. and {Castillo-Rogez}, Julie and {Nimmo}, Francis and {McEwen}, Alfred S. and {Gudipati}, Murthy S. and {Jun}, Insoo and {Jia}, Xianzhe and {Postberg}, Frank and {Soderlund}, Krista M. and {Elder}, Catherine M.},
        title = "{Investigating Europa's Habitability with the Europa Clipper}",
      journal = {\ssr},
         year = 2023,
        month = dec,
       volume = {219},
       number = {8},
          eid = {81},
        pages = {81},
          doi = {10.1007/s11214-023-01025-2},
       adsurl = {https://ui.adsabs.harvard.edu/abs/2023SSRv..219...81V}
}

@ARTICLE{Ki00,
       author = {{Kivelson}, Margaret G. and {Khurana}, Krishan K. and {Russell}, Christopher T. and {Volwerk}, Martin and {Walker}, Raymond J. and {Zimmer}, Christophe},
        title = "{Galileo Magnetometer Measurements: A Stronger Case for a Subsurface Ocean at Europa}",
      journal = {Science},
         year = 2000,
        month = aug,
       volume = {289},
       number = {5483},
        pages = {1340-1343},
          doi = {10.1126/science.289.5483.1340},
       adsurl = {https://ui.adsabs.harvard.edu/abs/2000Sci...289.1340K}
}

@ARTICLE{Bjonnes22,
       author = {{Bjonnes}, E. and {Johnson}, B.~C. and {Silber}, E.~A. and {Singer}, K.~N. and {Evans}, A.~J.},
        title = "{Ice Shell Structure of Ganymede and Callisto Based on Impact Crater Morphology}",
      journal = {Journal of Geophysical Research (Planets)},
         year = 2022,
        month = apr,
       volume = {127},
       number = {4},
          eid = {e07028},
        pages = {e07028},
          doi = {10.1029/2021JE007028},
       adsurl = {https://ui.adsabs.harvard.edu/abs/2022JGRE..12707028B}
}

@ARTICLE{Bennacer_2025,
       author = {{Bennacer}, Yannis and {Mousis}, Olivier and {Monnereau}, Marc and {Hue}, Vincent and {Schneeberger}, Antoine},
        title = "{Conditions for accretion favoring an unmelted Callisto and a differentiated Ganymede}",
      journal = {arXiv e-prints},
         year = 2025,
        month = may,
          eid = {arXiv:2505.07785},
        pages = {arXiv:2505.07785},
          doi = {10.48550/arXiv.2505.07785},
archivePrefix = {arXiv},
       eprint = {2505.07785},
 primaryClass = {astro-ph.EP},
       adsurl = {https://ui.adsabs.harvard.edu/abs/2025arXiv250507785B}
}

@ARTICLE{Petricca_2025,
       author = {{Petricca}, Flavio and {Castillo-Rogez}, Julie C. and {Genova}, Antonio and {Melwani Daswani}, Mohit and {Styczinski}, Marshall J. and {Cochrane}, Corey J. and {Vance}, Steven D.},
        title = "{Partial differentiation of Europa and implications for the origin of materials in the Jupiter system}",
      journal = {Nature Astronomy},
         year = 2025,
        month = apr,
       volume = {9},
        pages = {501-511},
          doi = {10.1038/s41550-024-02469-4},
       adsurl = {https://ui.adsabs.harvard.edu/abs/2025NatAs...9..501P}
}

@ARTICLE{Trinh_2023,
       author = {{Trinh}, Kevin T. and {Bierson}, Carver J. and {O'Rourke}, Joseph G.},
        title = "{Slow evolution of Europa's interior: metamorphic ocean origin, delayed metallic core formation, and limited seafloor volcanism}",
      journal = {Science Advances},
         year = 2023,
        month = jun,
       volume = {9},
       number = {24},
          eid = {eadf3955},
        pages = {eadf3955},
          doi = {10.1126/sciadv.adf3955},
       adsurl = {https://ui.adsabs.harvard.edu/abs/2023SciA....9F3955T}
}

@article{Anderson_2001,
title = {Shape, Mean Radius, Gravity Field, and Interior Structure of Callisto},
journal = {Icarus},
volume = {153},
number = {1},
pages = {157-161},
year = {2001},
issn = {0019-1035},
doi = {https://doi.org/10.1006/icar.2001.6664},
url = {https://www.sciencedirect.com/science/article/pii/S0019103501966643},
author = {J.D. Anderson and R.A. Jacobson and T.P. McElrath and W.B. Moore and G. Schubert and P.C. Thomas}
}

@ARTICLE{Barr_Canup_2008,
   author = {{Barr}, A.C and {Canup}, R.M},
        title = {Constraints on gas giant satellite formation from the interior states of partially differentiated satellites},
      journal = {Icarus},
         year = 2008,
        month = ...,
       volume = {198},
       number = {},
        pages = {163-177},
          doi = {},
archivePrefix = {arXiv},
       eprint = {},
 primaryClass = {},
       adsurl = {}
}

@article{Bierson_Nimmo_2020,
doi = {10.3847/2041-8213/aba11a},
url = {https://dx.doi.org/10.3847/2041-8213/aba11a},
year = {2020},
month = {jul},
publisher = {The American Astronomical Society},
volume = {897},
number = {2},
pages = {L43},
author = {Bierson, Carver J. and Nimmo, Francis},
title = {Explaining the Galilean Satellites‚Äô Density Gradient by Hydrodynamic Escape},
journal = {The Astrophysical Journal Letters}
}

@ARTICLE{Ochiai24,
       author = {{Ochiai}, Y. and {Ida}, S. and {Shoji}, D.},
        title = "{Monte Carlo simulation of UV-driven synthesis of complex organic molecules on icy grain surfaces}",
      journal = {\aap},
         year = 2024,
        month = jul,
       volume = {687},
          eid = {A232},
        pages = {A232},
          doi = {10.1051/0004-6361/202449655},
archivePrefix = {arXiv},
       eprint = {2406.00640},
 primaryClass = {astro-ph.EP},
       adsurl = {https://ui.adsabs.harvard.edu/abs/2024A&A...687A.232O}
}

@ARTICLE{Tak22,
       author = {{Takehara}, Hitoshi and {Shoji}, Daigo and {Ida}, Shigeru},
        title = "{Monte Carlo simulation of sugar synthesis on icy dust particles intermittently irradiated by UV in a protoplanetary disk}",
      journal = {\aap},
         year = 2022,
        month = jun,
       volume = {662},
          eid = {A76},
        pages = {A76},
          doi = {10.1051/0004-6361/202243212},
archivePrefix = {arXiv},
       eprint = {2203.06669},
 primaryClass = {astro-ph.EP},
       adsurl = {https://ui.adsabs.harvard.edu/abs/2022A&A...662A..76T}
}

@ARTICLE{Pr83,
       author = {{Prasad}, S.~S. and {Tarafdar}, S.~P.},
        title = "{UV radiation field inside dense clouds - Its possible existence and chemical implications}",
      journal = {\apj},
         year = 1983,
        month = apr,
       volume = {267},
        pages = {603-609},
          doi = {10.1086/160896},
       adsurl = {https://ui.adsabs.harvard.edu/abs/1983ApJ...267..603P}
}

@INPROCEEDINGS{Lin11,
       author = {{Linnartz}, Harold and {Bossa}, Jean-Baptiste and {Bouwman}, Jordy and {Cuppen}, Herma M. and {Cuylle}, Steven H. and {van Dishoeck}, Ewine F. and {Fayolle}, Edith C. and {Fedoseev}, Gleb and {Fuchs}, Guido W. and {Ioppolo}, Sergio and {Isokoski}, Karoliina and {Lamberts}, Thanja and {{\"O}berg}, Karin I. and {Romanzin}, Claire and {Tenenbaum}, Emily and {Zhen}, Junfeng},
        title = "{Solid State Pathways towards Molecular Complexity in Space}",
    booktitle = {The Molecular Universe},
         year = 2011,
       editor = {{Cernicharo}, Jos{\'e} and {Bachiller}, Rafael},
       series = {IAU Symposium},
       volume = {280},
        month = dec,
        pages = {390-404},
          doi = {10.1017/S1743921311025142},
       adsurl = {https://ui.adsabs.harvard.edu/abs/2011IAUS..280..390L}
}

@ARTICLE{Lich16,
       author = {{Lichtenberg}, Tim and {Parker}, Richard J. and {Meyer}, Michael R.},
        title = "{Isotopic enrichment of forming planetary systems from supernova pollution}",
      journal = {\mnras},
         year = 2016,
        month = nov,
       volume = {462},
       number = {4},
        pages = {3979-3992},
          doi = {10.1093/mnras/stw1929},
archivePrefix = {arXiv},
       eprint = {1608.01435},
 primaryClass = {astro-ph.EP},
       adsurl = {https://ui.adsabs.harvard.edu/abs/2016MNRAS.462.3979L}
}

@ARTICLE{Fat08,
       author = {{Fatuzzo}, Marco and {Adams}, Fred C.},
        title = "{UV Radiation Fields Produced by Young Embedded Star Clusters}",
      journal = {\apj},
         year = 2008,
        month = mar,
       volume = {675},
       number = {2},
        pages = {1361-1374},
          doi = {10.1086/527469},
archivePrefix = {arXiv},
       eprint = {0712.3487},
 primaryClass = {astro-ph},
       adsurl = {https://ui.adsabs.harvard.edu/abs/2008ApJ...675.1361F}
}

@ARTICLE{Adams04,
       author = {{Adams}, Fred C. and {Hollenbach}, David and {Laughlin}, Gregory and {Gorti}, Uma},
        title = "{Photoevaporation of Circumstellar Disks Due to External Far-Ultraviolet Radiation in Stellar Aggregates}",
      journal = {\apj},
         year = 2004,
        month = aug,
       volume = {611},
       number = {1},
        pages = {360-379},
          doi = {10.1086/421989},
archivePrefix = {arXiv},
       eprint = {astro-ph/0404383},
 primaryClass = {astro-ph},
       adsurl = {https://ui.adsabs.harvard.edu/abs/2004ApJ...611..360A}
}

@ARTICLE{Mousis2009,
       author = {{Mousis}, Olivier and {Marboeuf}, Ulysse and {Lunine}, Jonathan I. and {Alibert}, Yann and {Fletcher}, Leigh N. and {Orton}, Glenn S. and {Pauzat}, Fran{\c{c}}oise and {Ellinger}, Yves},
        title = "{Determination of the Minimum Masses of Heavy Elements in the Envelopes of Jupiter and Saturn}",
      journal = {\apj},
         year = 2009,
        month = may,
       volume = {696},
       number = {2},
        pages = {1348-1354},
          doi = {10.1088/0004-637X/696/2/1348},
archivePrefix = {arXiv},
       eprint = {0812.2441},
 primaryClass = {astro-ph},
       adsurl = {https://ui.adsabs.harvard.edu/abs/2009ApJ...696.1348M}
}

@INPROCEEDINGS{Benest_Couzinou_2025,
       author = {{Benest Couzinou}, Tom and {Amsler Moulanier}, Aliz{\'e}e. and {Mousis}, Olivier},
        title = "{Delivery of organic matter to the Galilean moons}",
    booktitle = {EGU General Assembly Conference Abstracts},
         year = 2025,
       series = {EGU General Assembly Conference Abstracts},
        month = apr,
          eid = {5889},
        pages = {5889},
          doi = {10.5194/egusphere-egu24-5889},
       adsurl = {https://ui.adsabs.harvard.edu/abs/2024EGUGA..26.5889B}
}

@ARTICLE{Schneeberger2024,
       author = {Schneeberger, Antoine and Mousis, Olivier},
        title = ".",
      journal = { Planet. Sci. J, Submitted},
         year = {2024},
        month = dec,
       volume = {},
       number = {E12},
          eid = {},
        pages = {},
          doi = {},
       adsurl = {}
}

@article{Bell_Lin_1994, title={Using FU Orionis Outbursts to Constrain Self-regulated Protostellar Disk Models}, volume={427}, ISSN={0004-637X}, DOI={10.1086/174206}, journal={The Astrophysical Journal}, author={Bell, K. R. and Lin, D. N. C.}, year={1994}, month=jun, pages={987} }

@ARTICLE{Ben24,
    author = {{Benest Couzinou}, T. and {Mousis}, O. and {Danger}, G. and {Schneeberger}, A. and {Aguichine}, A. and {Bouquet}, A.},
        title = "{Journey of complex organic molecules: Formation and transport in protoplanetary disks}",
      journal = {\aap},
         year = 2024,
        month = dec,
       volume = {692},
          eid = {A10},
        pages = {A10},
          doi = {10.1051/0004-6361/202449499},
archivePrefix = {arXiv},
       eprint = {2412.09271},
 primaryClass = {astro-ph.EP},
       adsurl = {https://ui.adsabs.harvard.edu/abs/2024A&A...692A..10B}
}

@article{Birnstiel2012, title={A simple model for the evolution of the dust population in protoplanetary disks}, volume={539}, ISSN={0004-6361}, DOI={10.1051/0004-6361/201118136}, note={ADS Bibcode: 2012A&A...539A.148B}, journal={Astronomy and Astrophysics}, author={Birnstiel, T. and Klahr, H. and Ercolano, B.}, year={2012}, month=mar, pages={A148} }

@article{bossa2008,
	title = {Carbamic acid and carbamate formation in {NH}\$\_\{{\textbackslash}sf 3\}\$:{CO}\$\_\{{\textbackslash}sf 2\}\$ ices – {UV} irradiation versus thermal processes},
	volume = {492},
	issn = {0004-6361, 1432-0746},
	shorttitle = {Carbamic acid and carbamate formation in {NH}\$\_\{{\textbackslash}sf 3\}\$},
	url = {http://www.aanda.org/10.1051/0004-6361:200810536},
	doi = {10.1051/0004-6361:200810536},
	language = {en},
	number = {3},
	urldate = {2024-01-08},
	journal = {Astronomy \& Astrophysics},
	author = {Bossa, J. B. and Theulé, P. and Duvernay, F. and Borget, F. and Chiavassa, T.},
	month = dec,
	year = {2008},
	pages = {719--724},
}

@ARTICLE{Cies11,
       author = {{Ciesla}, F.~J.},
        title = "{Residence Times of Particles in Diffusive Protoplanetary Disk Environments. II. Radial Motions and Applications to Dust Annealing}",
      journal = {\apj},
         year = 2011,
        month = oct,
       volume = {740},
       number = {1},
          eid = {9},
        pages = {9},
          doi = {10.1088/0004-637X/740/1/9},
archivePrefix = {arXiv},
       eprint = {1108.4736},
 primaryClass = {astro-ph.EP},
       adsurl = {https://ui.adsabs.harvard.edu/abs/2011ApJ...740....9C}
}

@ARTICLE{Cies10,
       author = {{Ciesla}, F.~J.},
        title = "{Residence Times of Particles in Diffusive Protoplanetary Disk Environments. I. Vertical Motions}",
      journal = {\apj},
         year = 2010,
        month = nov,
       volume = {723},
       number = {1},
        pages = {514-529},
          doi = {10.1088/0004-637X/723/1/514},
archivePrefix = {arXiv},
       eprint = {1010.1579},
 primaryClass = {astro-ph.EP},
       adsurl = {https://ui.adsabs.harvard.edu/abs/2010ApJ...723..514C}
}

@article{Ciesla_Sandford_2012, title={Organic Synthesis via Irradiation and Warming of Ice Grains in the Solar Nebula}, volume={336}, ISSN={0036-8075}, DOI={10.1126/science.1217291}, note={ADS Bibcode: 2012Sci...336..452C}, journal={Science}, author={Ciesla, Fred J. and Sandford, Scott A.}, year={2012}, month=apr, pages={452} }

@book{Cuzzi_Weidenschilling_2006, title={Particle-Gas Dynamics and Primary Accretion}, url={https://ui.adsabs.harvard.edu/abs/2006mess.book..353C}, note={ADS Bibcode: 2006mess.book..353C}, journal={Meteorites and the Early Solar System II}, author={Cuzzi, J. N. and Weidenschilling, S. J.}, publisher ={}, year={2006}, month=jan, pages={353} }

@article{Du95, title={The dust subdisk in the protoplanetary nebula.}, volume={114}, ISSN={0019-1035}, DOI={10.1006/icar.1995.1058}, note={ADS Bibcode: 1995Icar..114..237D}, journal={Icarus}, author={Dubrulle, B. and Morfill, G. and Sterzik, M.}, year={1995}, month=apr, pages={237–246} }

@article{Gail2001, title={Radial mixing in protoplanetary accretion disks. I. Stationary disc models with annealing and carbon combustion}, volume={378}, ISSN={0004-6361}, DOI={10.1051/0004-6361:20011130}, note={ADS Bibcode: 2001A&A...378..192G}, journal={Astronomy and Astrophysics}, author={Gail, H. -P.}, year={2001}, month=oct, pages={192–213} }

@article{Perets2011, title={Wind-shearing in Gaseous Protoplanetary Disks and the Evolution of Binary Planetesimals}, volume={733}, ISSN={0004-637X}, DOI={10.1088/0004-637X/733/1/56},  note={ADS Bibcode: 2011ApJ...733...56P}, journal={The Astrophysical Journal}, author={Perets, Hagai B. and Murray-Clay, Ruth A.}, year={2011}, month=may, pages={56} }

@article{tenelanda2022,
	title = {Effect of the {UV} dose on the formation of complex organic molecules in astrophysical ices: irradiation of methanol ices at 20 {K} and 80 {K}},
	volume = {515},
	issn = {0035-8711},
	shorttitle = {Effect of the {UV} dose on the formation of complex organic molecules in astrophysical ices},
	url = {https://ui.adsabs.harvard.edu/abs/2022MNRAS.515.5009T},
	doi = {10.1093/mnras/stac1932},
	urldate = {2023-04-26},
	journal = {Monthly Notices of the Royal Astronomical Society},
	author = {Tenelanda-Osorio, Laura I. and Bouquet, Alexis and Javelle, Thomas and Mousis, Olivier and Duvernay, Fabrice and Danger, Grégoire},
	month = oct,
	year = {2022},
	note = {ADS Bibcode: 2022MNRAS.515.5009T},
	pages = {5009--5017}
}

@article{Visser1997, 
    title={Using random walk models to simulate the vertical distribution of particles in a turbulent water column}, 
    volume={158}, 
    DOI={10.3354/meps158275}, 
    note={ADS Bibcode: 1997MEPS..158..275V}, 
    journal={Marine Ecology Progress Series}, 
    author={Visser, AW}, 
    year={1997}, 
    month=jan, 
    pages={275–281} }

@article{tanigawa2012,
  title = {Distribution of {{Accreting Gas}} and {{Angular Momentum}} onto {{Circumplanetary Disks}}},
  author = {Tanigawa, Takayuki and Ohtsuki, Keiji and Machida, Masahiro N.},
  year = {2012},
  month = mar,
  journal = {The Astrophysical Journal},
  volume = {747},
  pages = {47},
  publisher = {IOP},
  issn = {0004-637X},
  doi = {10.1088/0004-637X/747/1/47}
}

@article{lynden-bell1974,
  title = {The {{Evolution}} of {{Viscous Discs}} and the {{Origin}} of the {{Nebular Variables}}},
  author = {{Lynden-Bell}, D. and Pringle, J. E.},
  year = {1974},
  month = sep,
  journal = {Monthly Notices of the Royal Astronomical Society},
  volume = {168},
  number = {3},
  pages = {603--637},
  issn = {0035-8711, 1365-2966},
  doi = {10.1093/mnras/168.3.603}
}

@article{aguichine2022,
  title = {The {{Possible Formation}} of {{Jupiter}} from {{Supersolar Gas}}},
  author = {Aguichine, Artyom and Mousis, Olivier and Lunine, Jonathan I.},
  year = {2022},
  month = jun,
  journal = {The Planetary Science Journal},
  volume = {3},
  number = {6},
  pages = {141},
  issn = {2632-3338},
  doi = {10.3847/PSJ/ac6bf1}
}

@article{schneeberger2023,
  title = {Evolution of the Reservoirs of Volatiles in the Protosolar Nebula},
  author = {Schneeberger, Antoine and Mousis, Olivier and Aguichine, Artyom and Lunine, Jonathan I.},
  year = {2023},
  month = jan,
  journal = {Astronomy \& Astrophysics},
  eprint = {2301.02482},
  primaryclass = {astro-ph},
  issn = {0004-6361, 1432-0746},
  doi = {10.1051/0004-6361/202244670},
  archiveprefix = {arxiv}
}

@article{morbidelli2014,
       author = {{Morbidelli}, A. and {Szul{\'a}gyi}, J. and {Crida}, A. and {Lega}, E. and {Bitsch}, B. and {Tanigawa}, T. and {Kanagawa}, K.},
        title = "{Meridional circulation of gas into gaps opened by giant planets in three-dimensional low-viscosity disks}",
      journal = {\icarus},
         year = 2014,
        month = apr,
       volume = {232},
        pages = {266-270},
          doi = {10.1016/j.icarus.2014.01.010},
archivePrefix = {arXiv},
       eprint = {1401.2925},
 primaryClass = {astro-ph.EP},
       adsurl = {https://ui.adsabs.harvard.edu/abs/2014Icar..232..266M}
}

@article{zhu2012,
  title = {Challenges {{In Forming Planets By Gravitational Instability}}: {{Disk Irradiation And Clump Migration}}, {{Accretion}}, {{And Tidal Destruction}}},
  author = {Zhu, Zhaohuan and Hartmann, Lee and Nelson, Richard P. and Gammie, Charles F.},
  date = {2012-02-10},
  year = {2012},
  journal = {The Astrophysical Journal},
  shortjournal = {ApJ},
  volume = {746},
  number = {1},
  pages = {110},
  issn = {0004-637X, 1538-4357},
  doi = {10.1088/0004-637X/746/1/110}
}

@article{villenave2022,
       author = {{Villenave}, M. and {Stapelfeldt}, K.~R. and {Duch{\^e}ne}, G. and {M{\'e}nard}, F. and {Lambrechts}, M. and {Sierra}, A. and {Flores}, C. and {Dent}, W.~R.~F. and {Wolff}, S. and {Ribas}, {\'A}. and {Benisty}, M. and {Cuello}, N. and {Pinte}, C.},
        title = "{A Highly Settled Disk around Oph163131}",
      journal = {The Astrophysical Jouranal},
         year = 2022,
        month = may,
       volume = {930},
       number = {1},
          eid = {11},
        pages = {11},
          doi = {10.3847/1538-4357/ac5fae},
archivePrefix = {arXiv},
       eprint = {2204.00640},
 primaryClass = {astro-ph.SR},
       adsurl = {https://ui.adsabs.harvard.edu/abs/2022ApJ...930...11V}
}

@article{Lodders2019,
  title={Solar Elemental Abundances},
  author={Katharina Lodders},
  journal={Oxford Research Encyclopedia of Planetary Science},
  year={2019},
  url={https://api.semanticscholar.org/CorpusID:208527688}
}

@article{aguichine2020,
       author = {{Aguichine}, Artyom and {Mousis}, Olivier and {Devouard}, Bertrand and {Ronnet}, Thomas},
        title = "{Rocklines as Cradles for Refractory Solids in the Protosolar Nebula}",
      journal =  {The Astrophysical Journal},
         year = 2020,
        month = oct,
       volume = {901},
       number = {2},
          eid = {97},
        pages = {97},
          doi = {10.3847/1538-4357/abaf47},
archivePrefix = {arXiv},
       eprint = {2005.14116},
 primaryClass = {astro-ph.EP},
       adsurl = {https://ui.adsabs.harvard.edu/abs/2020ApJ...901...97A}
}

@article{lambrechts2012,
       author = {{Lambrechts}, M. and {Johansen}, A.},
        title = "{Rapid growth of gas-giant cores by pebble accretion}",
      journal = {\aap},
         year = 2012,
        month = aug,
       volume = {544},
          eid = {A32},
        pages = {A32},
          doi = {10.1051/0004-6361/201219127},
archivePrefix = {arXiv},
       eprint = {1205.3030},
}

@article{anderson2021,
  title = {Formation {{Conditions}} of {{Titan}}'s and {{Enceladus}}'s {{Building Blocks}} in {{Saturn}}'s {{Circumplanetary Disk}}},
  author = {Anderson, Sarah E. and Mousis, Olivier and Ronnet, Thomas},
  year = {2021},
  month = apr,
  journal = {The Planetary Science Journal},
  volume = {2},
  number = {2},
  pages = {50},
  issn = {2632-3338},
  doi = {10.3847/PSJ/abe0ba}
}

@article{canup2002,
  title = {Formation of the {{Galilean Satellites}}: {{Conditions}} of {{Accretion}}},
  shorttitle = {Formation of the {{Galilean Satellites}}},
  author = {Canup, Robin M. and Ward, William R.},
  year = {2002},
  month = dec,
  journal = {The Astronomical Journal},
  volume = {124},
  number = {6},
  pages = {3404--3423},
  issn = {00046256, 15383881},
  doi = {10.1086/344684}
}

@article{canup2006,
  title = {A Common Mass Scaling for Satellite Systems of Gaseous Planets},
  author = {Canup, Robin M. and Ward, William R.},
  year = {2006},
  month = jun,
  journal = {Nature},
  volume = {441},
  number = {7095},
  pages = {834--839},
  issn = {0028-0836, 1476-4687},
  doi = {10.1038/nature04860}
}

@article{heller2015,
       author = {{Heller}, Ren{\'e} and {Pudritz}, Ralph},
        title = "{Water Ice Lines and the Formation of Giant Moons around Super-Jovian Planets}",
      journal = {\apj},
         year = 2015,
        month = jun,
       volume = {806},
       number = {2},
          eid = {181},
        pages = {181},
          doi = {10.1088/0004-637X/806/2/181},
archivePrefix = {arXiv},
       eprint = {1410.5802},
 primaryClass = {astro-ph.EP},
       adsurl = {https://ui.adsabs.harvard.edu/abs/2015ApJ...806..181H}
}

@article{makalkin2014,
  title = {Accretion Disks around {{Jupiter}} and {{Saturn}} at the Stage of Regular Satellite Formation},
  author = {Makalkin, A. B. and Dorofeeva, V. A.},
  year = {2014},
  month = jan,
  journal = {Solar System Research},
  volume = {48},
  number = {1},
  pages = {62--78},
  issn = {0038-0946, 1608-3423},
  doi = {10.1134/S0038094614010067}
}

@article{mordasini2013,
  title = {Luminosity of Young {{Jupiters}} Revisited: {{Massive}} Cores Make Hot Planets},
  shorttitle = {Luminosity of Young {{Jupiters}} Revisited},
  author = {Mordasini, C.},
  year = {2013},
  month = oct,
  journal = {Astronomy \& Astrophysics},
  volume = {558},
  pages = {A113},
  issn = {0004-6361, 1432-0746},
  doi = {10.1051/0004-6361/201321617}
}

@article{Mousis2018, title={Synthesis of Molecular Oxygen via Irradiation of Ice Grains in the Protosolar Nebula}, volume={858}, ISSN={0004-637X}, DOI={10.3847/1538-4357/aab6b9}, note={ADS Bibcode: 2018ApJ...858...66M}, journal={The Astrophysical Journal}, author={Mousis, O. and Ronnet, T. and Lunine, J. I. and Maggiolo, R. and Wurz, P. and Danger, G. and Bouquet, A.}, year={2018}, month=may, pages={66} }

@article{Ronnet2017, title={Pebble Accretion at the Origin of Water in Europa}, volume={845}, ISSN={0004-637X}, DOI={10.3847/1538-4357/aa80e6}, note={ADS Bibcode: 2017ApJ...845...92R}, journal={The Astrophysical Journal}, author={Ronnet, Thomas and Mousis, Olivier and Vernazza, Pierre}, year={2017}, month=aug, pages={92} }

@article{sasaki2010,
  title = {Origin of the {{Different Architectures}} of the {{Jovian}} and {{Saturnian Satellite Systems}}},
  author = {Sasaki, T. and Stewart, G. R. and Ida, S.},
  year = {2010},
  month = may,
  journal = {The Astrophysical Journal},
  volume = {714},
  pages = {1052--1064},
  issn = {0004-637X},
  doi = {10.1088/0004-637X/714/2/1052}
}





\end{document}